\documentclass[11pt,a4paper,pdftex]{article}
\pdfoutput=1
\usepackage{jcappub}
\usepackage{color}

\newcommand{\be}{\begin{eqnarray}}
\newcommand{\ee}{\end{eqnarray}}

\title{A study for testing the Kerr metric with AGN iron line eclipses}

\author[a,b]{Alejandro~C\'ardenas-Avenda\~no,}
\author[a]{Jiachen~Jiang,}
\author[a,c,1]{Cosimo~Bambi,%
\note{Corresponding author}}

\affiliation[a]{Center for Field Theory and Particle Physics and Department of Physics,\\
Fudan University, 220 Handan Road, 200433 Shanghai, China}
\affiliation[b]{Programa de Matem\'atica, Fundaci\'on Universitaria Konrad Lorenz,\\ 
Carrera 9 Bis No. 62-43, 110231 Bogot\'a, Colombia}
\affiliation[c]{Theoretical Astrophysics, Eberhard-Karls Universit\"at T\"ubingen,\\ 
Auf der Morgenstelle 10, 72076 T\"ubingen, Germany}

\emailAdd{alejandro.cardenasa@konradlorenz.edu.co}
\emailAdd{jcjiang12@fudan.edu.cn}
\emailAdd{bambi@fudan.edu.cn}

\abstract{Recently, two of us have studied iron line reverberation mapping to test black hole candidates, showing that the time information in reverberation mapping can better constrain the Kerr metric than the time-integrated approach. Motivated by this finding, here we explore the constraining power of another time-dependent measurement: an AGN iron line eclipse. An obscuring cloud passes between the AGN and the distant observer, covering different parts of the accretion disk at different times. Similar to the reverberation measurement, an eclipse might help to better identify the relativistic effects affecting the X-ray photons. However, this is not what we find. In our study, we employ the Johannsen-Psaltis parametrisation, but we argue that our conclusions hold in a large class of non-Kerr metrics. We explain our results pointing out an important difference between reverberation and eclipse measurements.}


\begin{document}

\maketitle


\section{Introduction}

Over the past 60~years, general relativity has been tested in different regimes and environments. The Schwarzschild solution has been tested in the Solar System~\cite{gr1} and, recently, this quest has been also shifted to the full nonlinear regime of the theory by the Laser Interferometer Gravitational-wave Observatory (LIGO), which has demonstrated the existence of binary stellar-mass black hole systems by detecting for the first time gravitational waves produced by a binary black hole merger~\cite{gwligo}. It seems now that we are starting to test general relativity in the strong gravity regime, where deviations from standard predictions can more likely manifest. In this context, astrophysical black hole candidates appear to be the best laboratories to probe strong gravitational fields and to do so, there are at least two approaches. One is based on the study of their electromagnetic spectrum, mainly through the analysis of the features of the radiation emitted from the inner part of the accretion disk~\cite{rs,rl,rtj}. The second one relies on the study of gravitational waves radiated by systems with at least one black hole candidate~\cite{rgw}.

In the framework of general relativity, the spacetime around astrophysical black holes should be well described by the Kerr solution. Initial deviations from the Kerr metric are expected to be quickly radiated away with the emission of gravitational waves~\cite{k1}. The equilibrium electric charge can be reached very quickly, because of the highly ionised host environment of these objects, and it is completely negligible in the case of macroscopic bodies~\cite{k2}. Furthermore, the mass of the accretion disk is typically many orders of magnitude lower than the mass of the black hole and its impact on the geometry of the spacetime can be safely ignored~\cite{k3}. Therefore, the detection of possible deviations from the Kerr solution should thus be attributed to new physics.

The continuum-fitting and the iron line methods are currently the leading techniques to probe the spacetime geometry around black hole candidates~\cite{cfm1,cfm2,cfm3,fe1,fe2,fe3}, and their use to test the Kerr metric has been investigated in a number of recent papers~\cite{t1,t2,t3,t4,t4b,t5,t6,t7,t8,t9,jjc2,harko1,harko2,harko3,t10} (for a review, see Ref.~\cite{rev}). The main problem to test black hole candidates is parameter degeneracy: it is always challenging to constrain new physics because features in the spectrum associated to possible deviations from the Kerr geometry may be reproduced by standard physics with a different configuration of the system.

Recently, two of us have investigated in Refs.~\cite{jjc1,jjc3} the possibility of testing the Kerr metric with iron line reverberation mapping, i.e., the time-evolution of the iron line profile in response to fluctuations in the X-ray primary source. The results unambiguously show that the time information in reverberation mapping can better constrain the background metric than the time-integrated iron line measurement, and this is true even for deviations from the Kerr geometry that do not leave any characteristic feature in the time-integrated profile.

Motivated by such a positive result, here we study the constraining power of another time-dependent measurement, an AGN iron line eclipse, which is expected to be a not too rare event~\cite{ecl}\footnote{We have already evidence for AGN X-ray eclipses. The best studied case is NGC~1365~\cite{2e1}. Other examples are NGC~4388~\cite{2e2}, NGC~4151~\cite{2e3}, and NGC~7582~\cite{2e4}.}. In this context, an obscuring cloud passes between the AGN and the observer, covering different parts of the accretion disk at different times. Similar to the reverberation measurement of an AGN, the iron line eclipse may offer the opportunity to better identify the relativistic effects affecting the X-ray photons, which would help to better constrain the Kerr metric.

Our simulations, however, show that this is not the case. In the present work, we adopt the Johannsen-Psaltis parametrisation to quantify possible deviations from the Kerr background. While we do not have any proof that our conclusions hold for any kind of deviations from the Kerr metric, we expect that the result is very general. As we discuss in this paper, there are a few important differences between reverberation mapping and eclipse measurement. We find that the key-point is related to the capability of separating photons from different parts of the disk. In the reverberation approach, photons emitted from different regions are detected at different time, and this is a very clean way to study the relativistic effects from each patch of the disk. However, in the eclipse scenario we have the opposite case, namely we have to figure out the properties of the radiation from every region of the accretion disk from the non-detection of the photons from that patch. In other words, we have to figure out the properties of the radiation by subtracting different 
spectra. If the cloud covers a small region, it is difficult to measure a difference. If the cloud covers a large region, we lose the information of the exact emission region on the disk.

Other effects seem to play a minor role in the difference between reverberation mapping and eclipse measurement. For instance, in the eclipse measurement we have also a loss of photons due to the passage of the cloud, while no loss is in the reverberation observation with respect to the standard time-integrated measurement. This also does not improve an eclipse measurement.

This paper is organised as follows. The theoretical set-up of our work is presented in Section~\ref{sec:TF}. Section~\ref{sec:S} describes the set-up and methods of our simulations. Section~\ref{sec:D} is devoted to discuss our results and Section~\ref{sec:C} for the conclusions.

\section{Theoretical framework}\label{sec:TF}

The electromagnetic spectrum depends on the motion of the gas in the accretion disk and on the propagation of the photons from the emission point in the strong gravity field to the detection point in the flat faraway region. Tests based on the study of the electromagnetic radiation can thus probe the metric around a black hole candidate. For instance, with this approach we cannot distinguish a Kerr black hole of general relativity from a Kerr black hole in another theory of gravity, because there is no difference in the geodesic motion~\cite{pkk}. This is exactly the same situation as that of experiments of general relativity in the Solar System, in which we can test the Schwarzschild solution in the weak field limit, but we cannot distinguish the Schwarzschild metric in general relativity from the Schwarzschild metric in another theory of gravity.

The electromagnetic radiation emitted by the gas in the inner part of the accretion disk is strongly affected by the relativistic effects experienced by these photons when they are close to the black hole candidate. Therefore, the geometry of the spacetime produces specific signatures in the electromagnetic spectrum, allowing us, at least in principle, to test the nature of the compact object.

In order to test the Kerr metric in a model-independent way and quantify possible deviations from the Kerr solution, it is common to adopt an approach that reminds the Parametrised Post-Newtonian (PPN) formalism, which is used to test the Schwarzschild solution in the weak field limit in the Solar System. In the PPN formalism, the starting point is to assume the most general static, spherically symmetric, and asymptotically flat line element. Since we are in the weak field limit, we can perform an expansion in $M/r \ll 1$. The line element in isotropic coordinates reads
\be
ds^2 = - \left( 1 - \frac{2M}{r} + \beta \frac{2M^2}{r^2} + . . . \right) dt^2
+ \left( 1 + \gamma \frac{2M}{r} + . . . \right) \left( dx^2 + dy^2 + dz^2 \right) \, ,
\ee
where $\beta$ and $\gamma$ are coefficients that parametrise our ignorance and must be obtained from observations. The only spherically symmetric vacuum solution of the Einstein equation is the Schwarzschild metric, where $\beta = \gamma = 1$. Current Solar System experiments constrain $\beta$ and $\gamma$ to be 1 with a precision, respectively, of $10^{-4}$ and $10^{-5}$~\cite{gr1}. This confirms the Schwarzschild solution at this level of accuracy.

In the case of black holes, we consider a metric that is more general than the Kerr solution and that includes the Kerr solution as a special case. Such a metric will be characterised by the mass $M$ and the spin angular momentum $J$ of the compact object, as well as by a number of ``deformation parameters''~\cite{met1,jp-m,met2,met3,met4,met5}. The latter are used to quantify possible deviations from the Kerr geometry and are assumed {\it a priori} unknown constants to be determined by observations. In other words, by measuring these deformation parameters we can check whether they vanish, as it is required by the Kerr metric. If observations required at least a non-vanishing deformation parameter, black hole candidates would not be the Kerr black holes of general relativity.

Due to the nature of the simulations we do in this work, in which we range over several parameters, described in Sec.~\ref{sec:S}, we have employed the simplest version of the Johannsen-Psaltis metric presented in Ref.~\cite{jp-m}, in which there is only one deformation parameter $\epsilon_3$. In Boyer-Lindquist coordinates, the line element reads
\be
ds^2 &=& - \left(1 - \frac{2 M r}{\Sigma}\right)\left(1 + h\right) dt^2
 - \frac{4 M a r \sin^2\theta}{\Sigma}\left(1 + h\right) dt d\phi 
+ \frac{\Sigma \left(1 + h\right)}{\Delta + h a^2 \sin^2\theta} dr^2
+ \Sigma d\theta^2 \nonumber\\
&& + \left[r^2 + a^2 + \frac{2 a^2 M r \sin^2\theta}{\Sigma}
+ \frac{a^2 \left(\Sigma + 2 M r\right) \sin^2\theta}{\Sigma} h \right] 
\sin^2\theta d\phi^2 \, ,
\ee
where $a=J/M$ is the specific spin, $\Sigma = r^2 + a^2 \cos^2\theta$, $\Delta = r^2 - 2 M r + a^2$, and 
\be
h &=& \epsilon_3 \frac{M^3 r}{\Sigma^2} \, .
\ee
If $\epsilon_3 = 0$, the object is a Kerr black hole. If $\epsilon_3 > 0$ ($< 0$), the object is more oblate (prolate) than a Kerr black hole with the same spin parameter~\cite{obl-prol}.

We have to note that the Johannsen-Psaltis metric is quite artificial. It is not a solution of any known theory of gravity. It is just a black hole metric in which the deformation parameter is introduced by hand and the Kerr metric is recovered when the deformation parameter vanishes. Despite that, it is commonly used as a test-metric for this kind of studies. First, it is usually extremely difficult to find an exact rotating black hole solution in an alternative theory of gravity, so the choice of a phenomenological metric is the simplest option. Second, as shown by previous work, the choice of the parametrisation is usually not very important for a simple analysis as the one presented in this paper. Of course, some deviations from the Kerr metric are easier to constrain than others, but the qualitative results found within one parametrisation often hold in a much larger context. Bearing all these points in mind, in this paper we study tests of the Kerr metric within the Johannsen-Psaltis parametrisation and we expect that the qualitative conclusions are true for a large group of non-Kerr metrics.

\section{Set-up of the simulation}\label{sec:S}

Since the aim of this paper is to study the constraining power of an iron line eclipse measurement with respect to the standard time-integrated observation without eclipse, we have performed some simulations of iron line measurement either with or without eclipse, with the set-up described bellow.

We employ the Novikov-Thorne model~\cite{2ntm,2ntm2} (see e.g. Ref.~\cite{rev} for more details and the validity of this model). The accretion disk is in the plane perpendicular to the black hole spin. The particles of the gas follow nearly geodesic equatorial circular orbits. The inner edge of the accretion disk is at the radius of the innermost stable circular orbit (ISCO). When the particles of the gas reach the ISCO, they quickly plunge onto the central object, without emitting additional radiation. There are a number of assumptions behind this model, but eventually it should work for geometrically thin and optically thick accretion disk in which the accretion rate is between $\sim$5\% and $\sim$30\% the Eddington limit~\cite{rev}.

The background geometry is described by the dimensionless spin parameter $a_* = a/M$ and the deformation parameter $\epsilon_3$ of the metric introduced in Sec.~\ref{sec:TF}. The mass $M$ only sets the size of the system and it does not directly affects the reflected component. The viewing angle is $i$ and corresponds to the angle between the spin axis and the line of sight of the observer. For the sake of simplicity, the emissivity profile is taken to be a power law with emissivity index $q = 3$; that is, $I_{\rm e} \propto r^{-3}$. The iron line signal is added to a power-law continuum with photon index $\Gamma$.

The simulations for the time-integrated measurements are done as in Ref.~\cite{jjc2}. We consider a reference Kerr model with spin parameter $a_*'$, viewing angle $i'$, we assume a certain number of counts in the iron line, $N_{\rm line}'$, and we add the power-law continuum with photon index $\Gamma' = 2$ to have 100 times the number of iron line photons when integrated over the energy range 1-9~keV. This corresponds to an equivalent width of $EW \approx 370$-440~eV, depending on the line shape.

We add Poisson noise\footnote{Generally speaking, the Poisson noise is any kind of noise that can be modelled by a Poisson process. Here it is present because the detector counts photons and not a continuous quantity.} to the reference model and treat this spectrum as a real observation. The spectrum is binned to achieve threshold counts, here $n_{\rm min} = 20$, and is then compared with the spectra expected from a Johannsen-Psaltis black hole with spin parameter $a_*$, deformation parameter $\epsilon_3$, viewing angle $i$, photon index of the continuum $\Gamma$, and ratio between the continuum and the iron line flux $K$. The other parameters of the model are fixed.

Let us use the notation $n_k' = n_k'(a_*', i', \Gamma', K')$ and $n_k = n_k(a_*, \epsilon_3, i, \Gamma, K)$ to indicate, respectively, the photon flux number density in the energy bin $[E_k , E_k + \Delta E]$ of the reference Kerr model and of Johannsen-Psaltis model (in our simulations, $\Delta E = 0.1$~keV). The normalised (negative) likelihood is
\be
\mathcal{L} = \frac{1}{\sum_k n_k} \left[ \sum_k \frac{\left(n_k - \alpha n_k'\right)^2}{n_k} \right] \, ,
\ee
where $\alpha$ is chosen to minimise $\mathcal{L}$, and therefore
\be
\alpha = \frac{\sum_k n_k'}{\sum_k n_k'^2/n_k} \, .
\ee
The chi-square is $\chi^2 \approx N_{\rm tot} \mathcal{L}$ and we study the contour levels of $\Delta\chi^2 = \chi^2 - \chi^2_{\rm min}$ to find the constraints that could be obtained from a similar observation.

\begin{figure}
\begin{center}
\hspace{-1.4cm}
\includegraphics[type=pdf,ext=.pdf,read=.pdf,width=9.5cm]{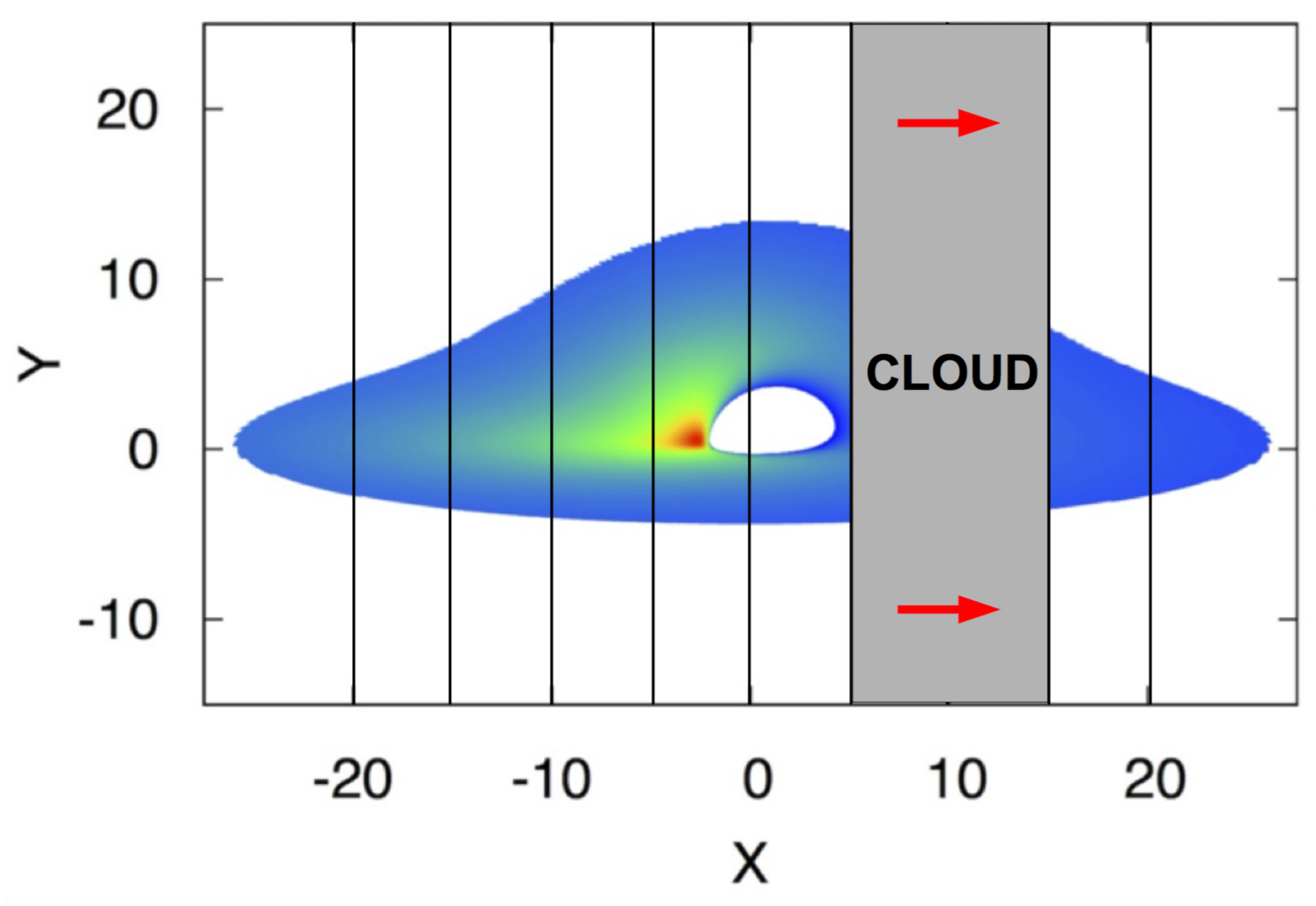}
\end{center}
\vspace{-0.4cm}
\caption{Sketch of our set-up to describe the observation of an AGN eclipse. In this picture, the image plane of the 
observer is divided into 8~slices, every slices has a width of $5 \, M$, and the cloud has a width of $10 \, M$ (X and 
Y 
in units in which $M=1$). See the text for more details.}
\label{f-image}
\end{figure}

For the eclipse scenario, the approach is the same and we consider a set of different configurations. Our general set-up is schematically illustrated in Fig.~\ref{f-image}. The image plane of the distant observer is divided into a number vertical slices, all of the same width. The obscuring cloud completely covers some slices (the exact number depends on the width of the cloud) and moves from one side of the image to the other side. If the photon count in the iron line without eclipse is $N_{\rm line}$, in the presence of an eclipse is lower. The exact number mainly depends on the size of the cloud and only weakly on the background metric, the viewing angle, and the slice width. In the simplest case, every measurement of the reflected component corresponds to a static configuration and the measurement after is that in which the cloud has moved by one slice. We have also studied more complicated configurations that takes into account the cloud motion and every measurement corresponds to a set of different static configurations.

Unlike reverberation observations studied in Ref.~\cite{jjc3}, where only the height of the corona matters and the behaviour of the system is otherwise established by the constant of the speed of light, here different eclipse configurations can provide different results. The number and width of slices and the width of the cloud inevitably affect the final result and the constraining capability of an eclipse measurement.

For instance, if the number of measurements increases, namely the number of observations in which the cloud is in different positions, the constraining power typically improves if we have sufficient photons in the iron line and it is irrelevant for a low number of photon count. This pattern is completely understandable, since more observations means a higher resolution in the tomography of the accretion disk. However, this also increases the number of channels and dilutes the photons. If the photon count is low, the intrinsic noise of the source prevents an improvement of the measurement and eventually the accuracy of an observation is determined by $N_{\rm line}$, namely by the effective area of the X-ray detector, and a higher time resolution does not provide any advantage.

The size of the cloud plays also a role in the final result. If the width of the cloud is too large, the photon count decreases, but even the power of the resolution of the tomography. If the cloud is too small, we need a very high number of photon count to be able to reconstruct the properties of the radiation in the region covered by the cloud.

\begin{figure}
\begin{center}
\includegraphics[type=pdf,ext=.pdf,read=.pdf,width=7.3cm]{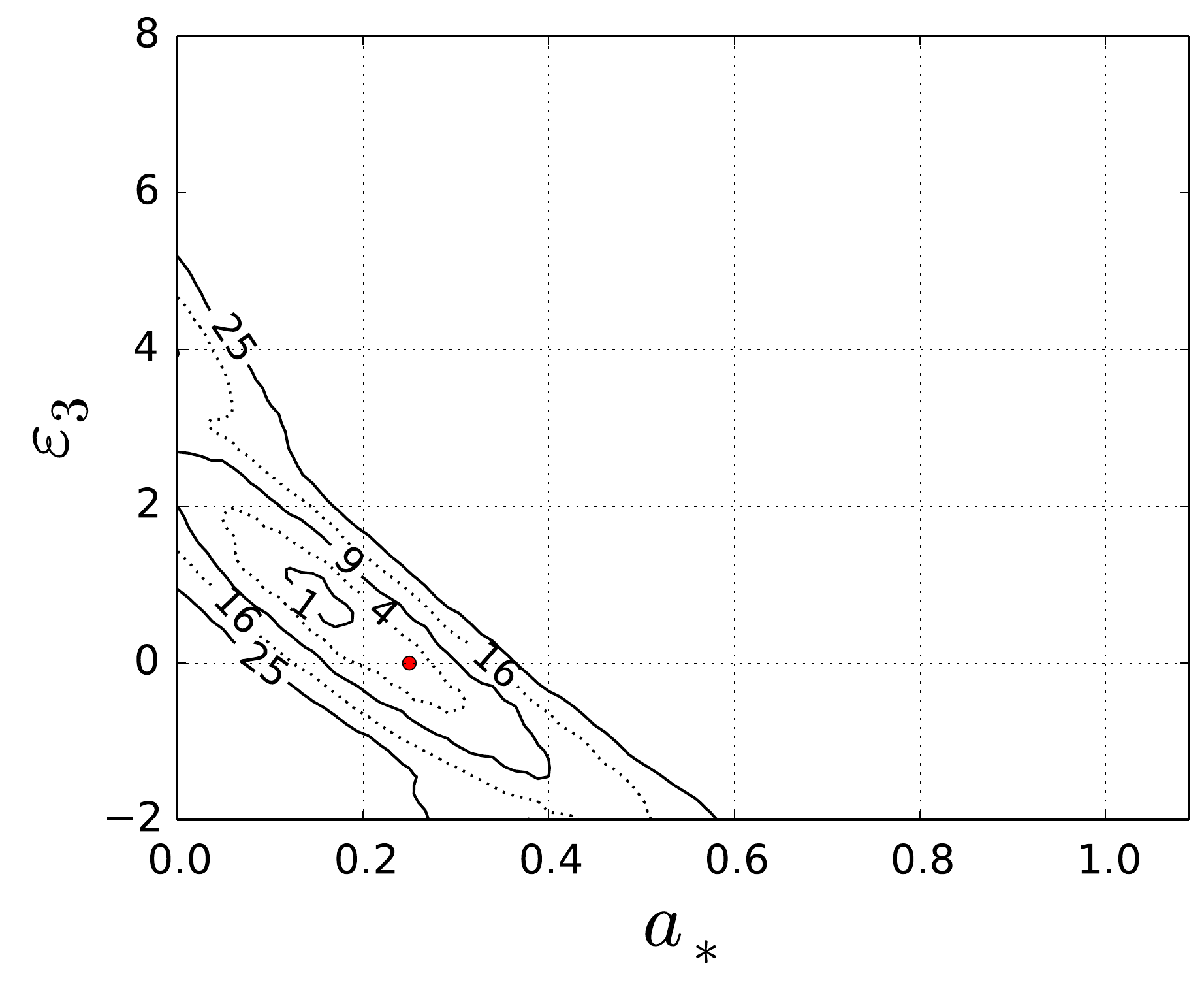}
\hspace{0.4cm}
\includegraphics[type=pdf,ext=.pdf,read=.pdf,width=7.3cm]{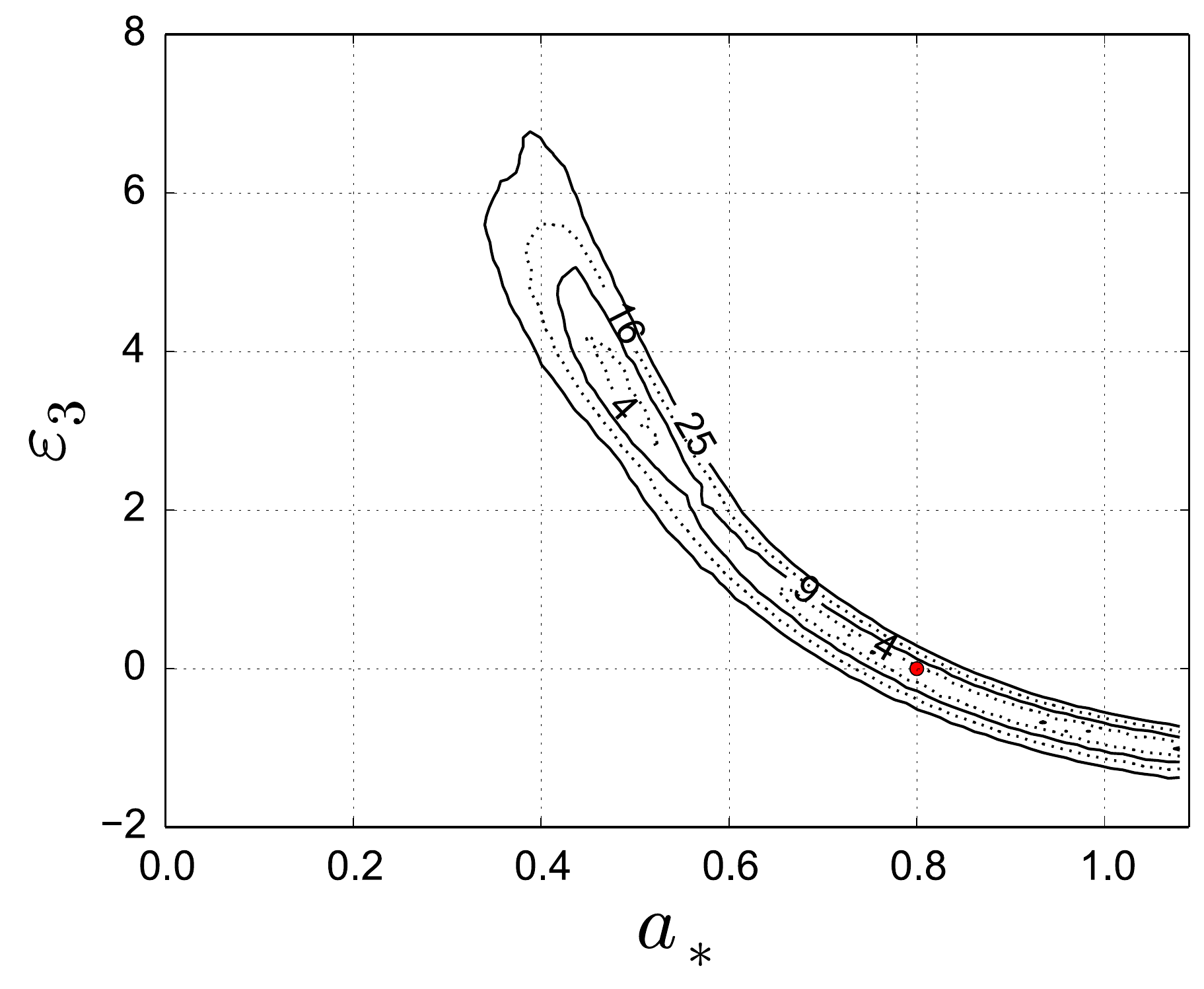}
\end{center}
\vspace{-0.4cm}
\caption{Standard time-integrated measurement without eclipse (model~0): $\Delta\chi^2$ contours with $N_{\rm line} = 
10^4$ from the comparison of the iron line of a Kerr black hole simulated using an input parameter $a_*' = 0.25$ (left 
panel) or $a_*' = 0.8$ (right panel) and a viewing angle $i' = 45^\circ$ vs a set of Johannsen-Psaltis black holes with 
spin parameter $a_*$, non-vanishing deformation parameter $\epsilon_3$, and arbitrary viewing angle $i$. See the text 
for more details. \label{fig1}} 
\vspace{0.6cm}
\begin{center}
\includegraphics[type=pdf,ext=.pdf,read=.pdf,width=7.3cm]{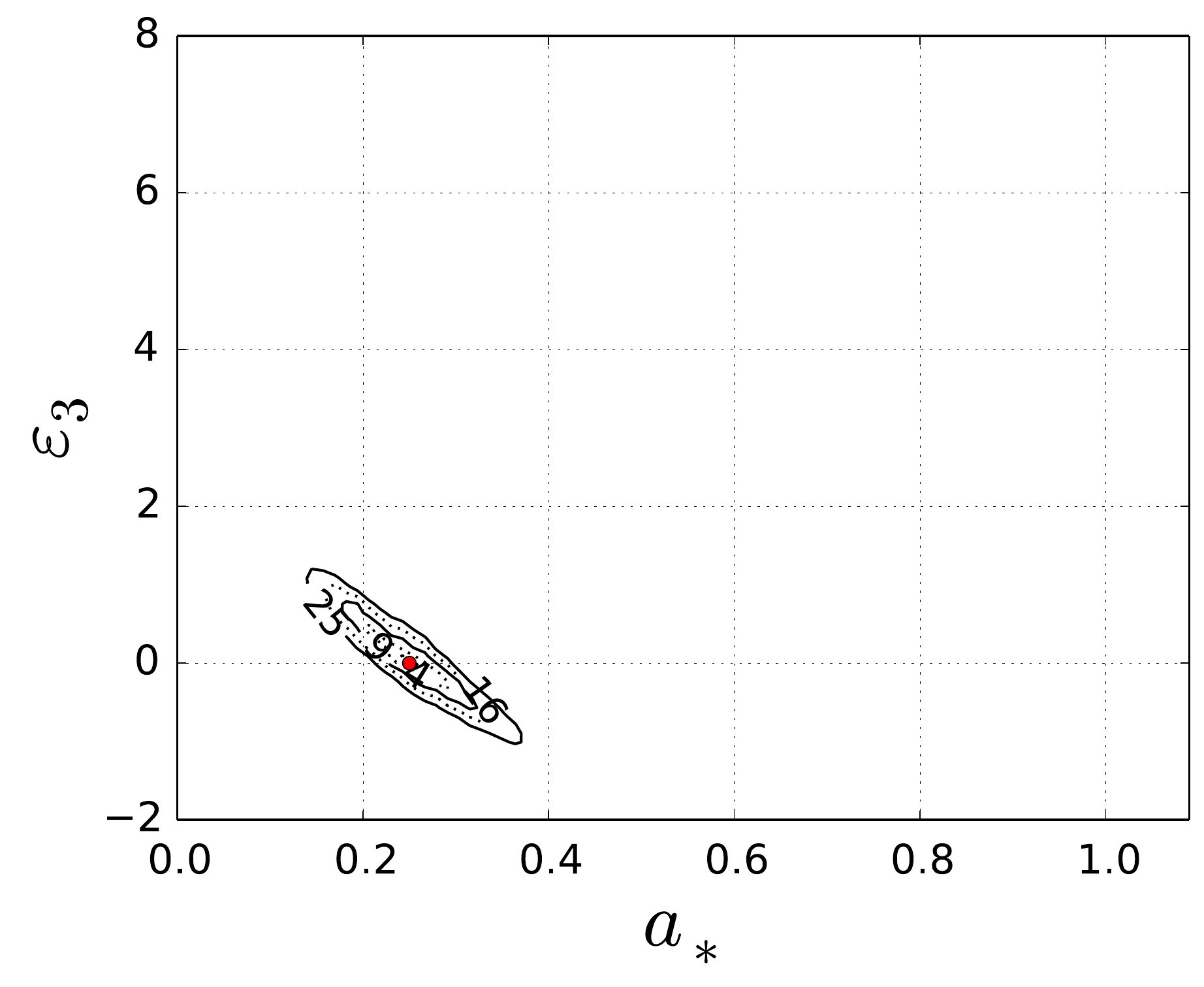}
\hspace{0.4cm}
\includegraphics[type=pdf,ext=.pdf,read=.pdf,width=7.3cm]{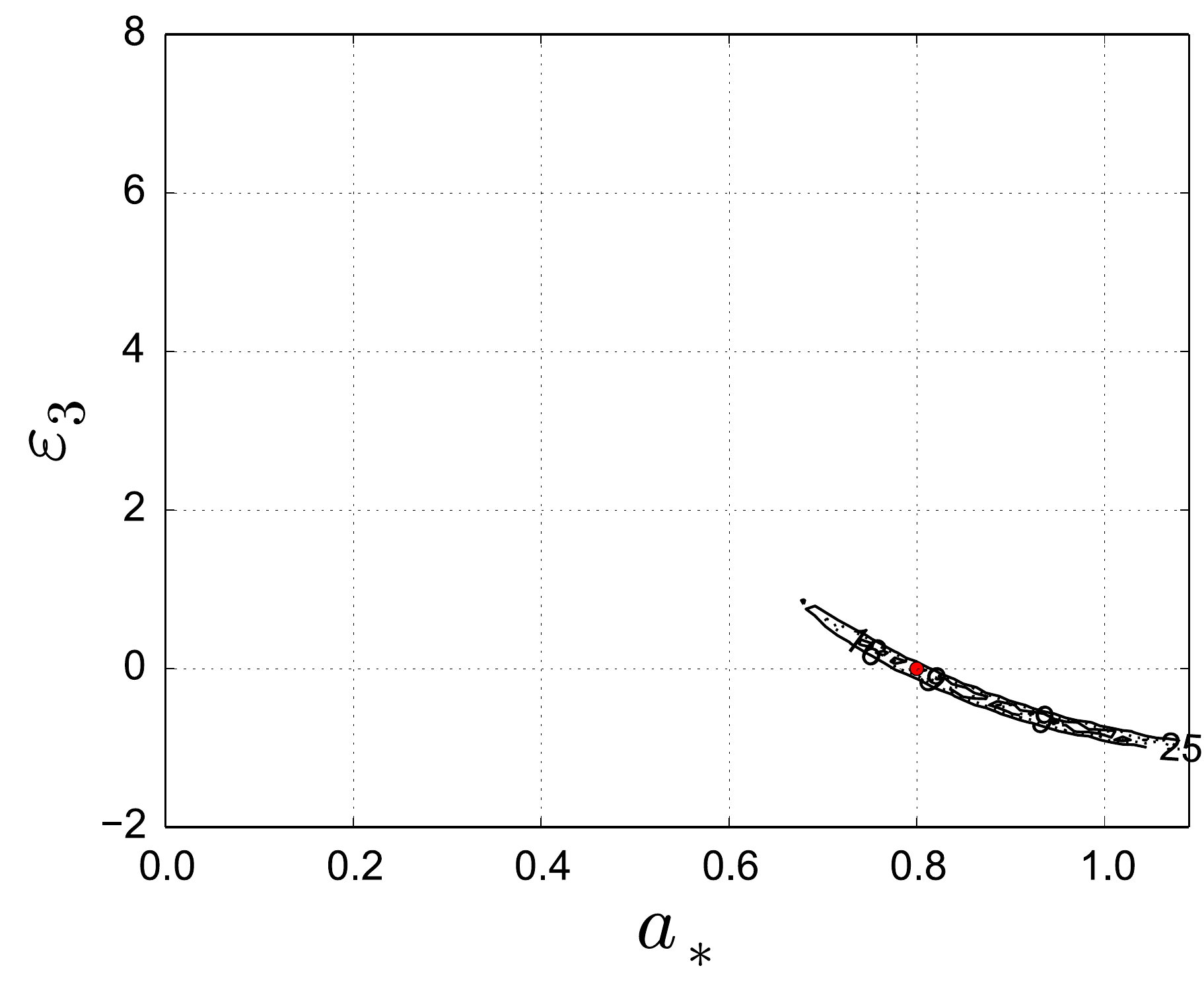}
\end{center}
\vspace{-0.4cm}
\caption{As in Fig.~\ref{fig1}, but for $N_{\rm line} = 10^5$. See the text for more details. \label{fig2}}
\end{figure}

We have considered a number of different configurations in the eclipse case, by changing the number of slices, the number of observations of the eclipse and the size and geometry of the obscuring cloud. We have also investigated the impact of different black hole spins and viewing angles. With our initial surprise, we have always found that the time-integrated and the eclipse measurements substantially provide similar results. Depending on the exact choice of the eclipse model and on the photon count in the iron line $N_{\rm line}$, the eclipse measurement can be somewhat better or somewhat worse than the standard time-integrated measurement. This is definitively in sharp contrast with the results for reverberation mapping~\cite{jjc3}. In the case of reverberation, its constraining power is already slightly better than the time-integrated observation for $N_{\rm line} = 10^3$, which roughly corresponds to a current good observation of an AGN. For $N_{\rm line} = 10^4$, the reverberation measurement is clearly better and it is also possible to constrain some deviations from Kerr that do not leave specific signatures in the time-integrated 
observations~\cite{jjc3}.

Figs.~\ref{fig1}-\ref{fig2} show the constraints obtained by simulating standard time-integrated measurements without eclipse. Fig.~\ref{fig1} is for $N_{\rm line} = 10^4$, while Fig.~\ref{fig2} shows the results for $N_{\rm line} = 10^5$. The reference Kerr black hole is indicated by the position of the red dot in the plane $(a_*, \epsilon_3)$. The spin parameter is $a_*' = 0.25$ in the left panels and $a_*' = 0.8$ in the right panels. As $N_{\rm line} $ increases, the effect of the Poisson noise is reduced, and the constraint gets stronger.

\begin{figure}
\begin{center}
\includegraphics[type=pdf,ext=.pdf,read=.pdf,width=7.3cm]{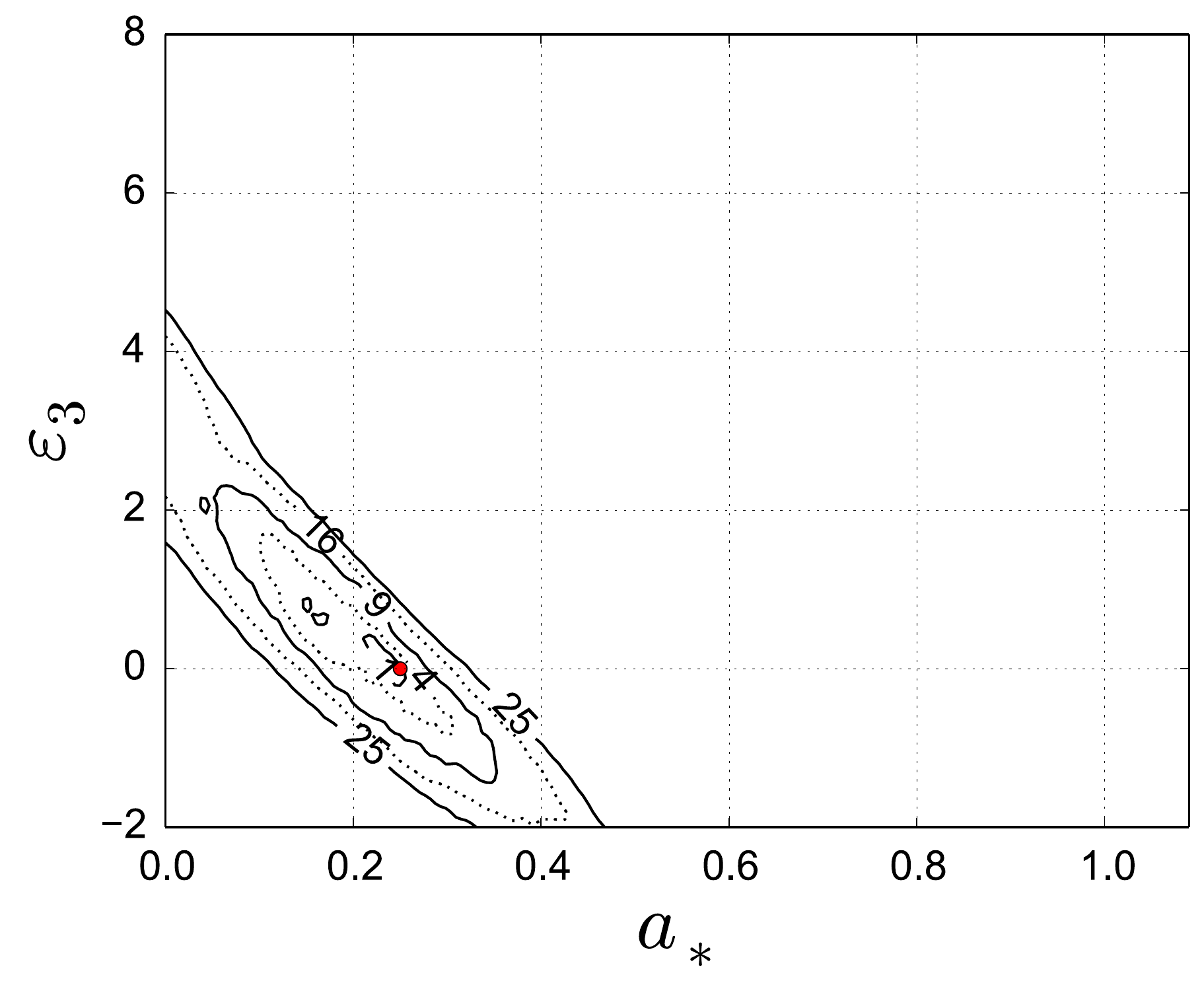}
\hspace{0.4cm}
\includegraphics[type=pdf,ext=.pdf,read=.pdf,width=7.3cm]{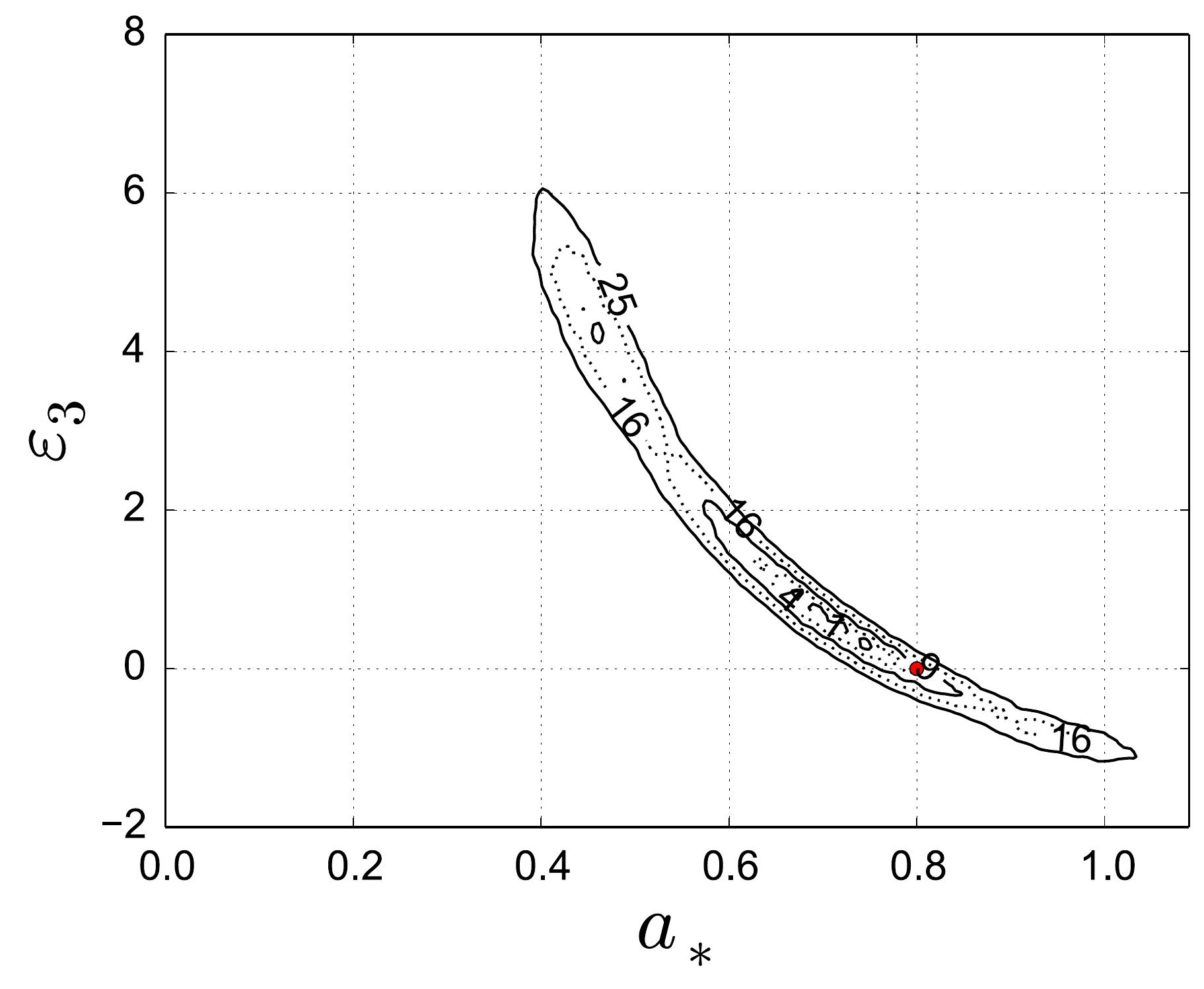}
\end{center}
\vspace{-0.4cm}
\caption{Eclipse measurement: $\Delta\chi^2$ contours with $N_{\rm line} = 10^4$ (when there is no eclipse) from the 
comparison of the iron line of a Kerr black hole simulated using an input parameter $a_*' = 0.25$ (left panel) or $a_*' 
= 0.8$ (right panel) and a viewing angle $i' = 45^\circ$ vs a set of Johannsen-Psaltis black holes with spin parameter 
$a_*$, non-vanishing deformation parameter $\epsilon_3$, and arbitrary viewing angle $i$. We have considered 
16~measurements and a cloud with width $6 \, M$. See the text for more details. \label{fig3}}
\end{figure}

Fig.~\ref{fig3} shows the constraints obtained by simulating an eclipse for $N_{\rm line} = 10^4$. As in Figs.~~\ref{fig1}-\ref{fig2}, the spin parameter is $a_*' = 0.25$ in the left panel and $a_*' = 0.8$ in the right panel. Here the slice width is $2 \, M$, the cloud width is $6 \, M$, and we consider 16~observations (at any observation, the cloud simply moves by one slice). The cloud starts from $X \approx - 15 \, M$ in the image plane of the observer and at the end of the observation is at $X \approx 15 \, M$. These plots clearly show that the standard approach and the eclipse measurement provide essentially the same kind of constraints. In the simulations of Ref.~\cite{jjc3}, the time-information in the reverberation measurement was a clear advantage for testing the Kerr metric and constraining the deformation parameter. The same constraining power of reverberation mapping with $N_{\rm line} = 10^4$ was already somewhat better than a time-integrated measurement with $N_{\rm line} = 10^5$.

\section{Discussion}\label{sec:D}

In the previous section, we have described our set-up and showed the results of some simulations. While we have considered several different configurations by changing the parameters of the eclipse model, our conclusion is that the time-information in the observation of an AGN eclipse does not provide the unambiguous advantages found in the reverberation case to test the Kerr metric. In this section, we try to understand the reason.

The simplest way to figure out the difference between a reverberation and an eclipse measurement is probably to consider a set of models, ranging from one very similar to a reverberation measurement, for which we expect to recover the reverberation results, to an eclipse model. With this spirit, we consider the following five models:
\begin{enumerate}
\item Model~0 -- Standard time-integrated iron line measurement, to be compared with the other 
models to see when and why the other models provide better results. We consider $N_{\rm line} = 10^4$ (photons in the iron line), and we adopt a small disk with inner edge at the ISCO and outer edge at the radius $r_{\rm out} = r_{\rm ISCO} + 16 \, M$. These choices will be adopted even in the other models.  
\item  Model~A -- For the model similar to a reverberation measurement, we divide the disk into 
16~regions, each of them is an annulus with an inner radius $r_{\rm in}$ and an outer radius $r_{\rm out} = r_{\rm in} + M$. The first annulus has $r_{\rm in} = r_{\rm ISCO}$ and $r_{\rm out} = r_{\rm ISCO} + M$. The second annulus has $r_{\rm in} = r_{\rm ISCO} + M$ and $r_{\rm out} = r_{\rm ISCO} + 2\,M$, etc. until the last annulus with $r_{\rm in} = r_{\rm ISCO} + 15\,M$ and $r_{\rm out} = r_{\rm ISCO} + 16\,M$. Similar to a reverberation measurement, we assume to be able to measure the iron line profile from each annulus. Then we compare every annulus of the reference model with its counterpart of the comparison model. 
\item Model~B -- In order to consider something between the model~A and an eclipse observation, we divide the image plane of the distant observer into 16~vertical slices. However, like in a reverberation measurement, we assume to be able to measure the 16~spectra from each slice and we proceed in the data analysis as in the case of the model~A, by comparing the spectrum from every slice of the reference model with its counterpart of the comparison model.
\item Model~C -- We start again from the model~A and we consider a modification to make it closer to an eclipse, but different from the model~B. We maintain the configurations with 16~annuli, but we assume that a ``cloud'' only covers one of this annulus. In the first observation, there is no cloud. In the second observation, the cloud covers the first annulus with $r_{\rm in} = r_{\rm ISCO}$ and $r_{\rm out} = r_{\rm ISCO} + M$. In the third observation, the cloud covers the second annulus with $r_{\rm in} = r_{\rm ISCO} + M$ and $r_{\rm out} = r_{\rm ISCO} + 2\,M$, and so on. The cloud moves to larger radii and in the last observation it covers the annulus with $r_{\rm in} = r_{\rm ISCO} + 14\,M$ and $r_{\rm out} = r_{\rm ISCO} + 15\,M$.  
\item Model~D -- This model corresponds to an eclipse measurement with 16~vertical slices. The cloud has a width of $6 \, M$ The model~D can be obtained from the model~B by introducing the modification between the model~A and the model~C. Alternatively, it can be obtained from the model~C by introducing the modification between the model~A and the model~B. 
\end{enumerate}

\begin{figure}
\begin{center}
\includegraphics[type=pdf,ext=.pdf,read=.pdf,width=7.0cm]{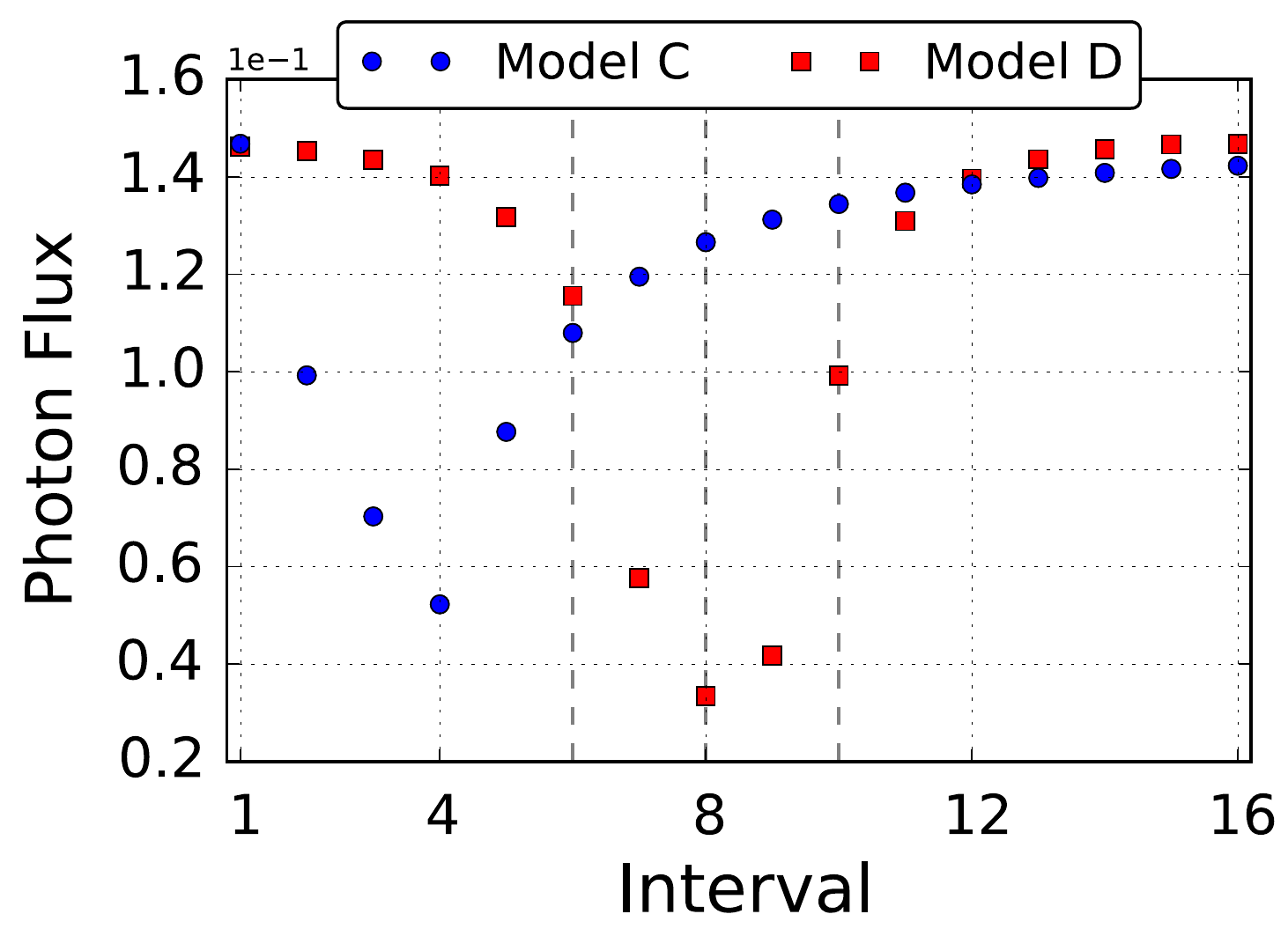}
\hspace{0.5cm}
\includegraphics[type=pdf,ext=.pdf,read=.pdf,width=7.3cm]{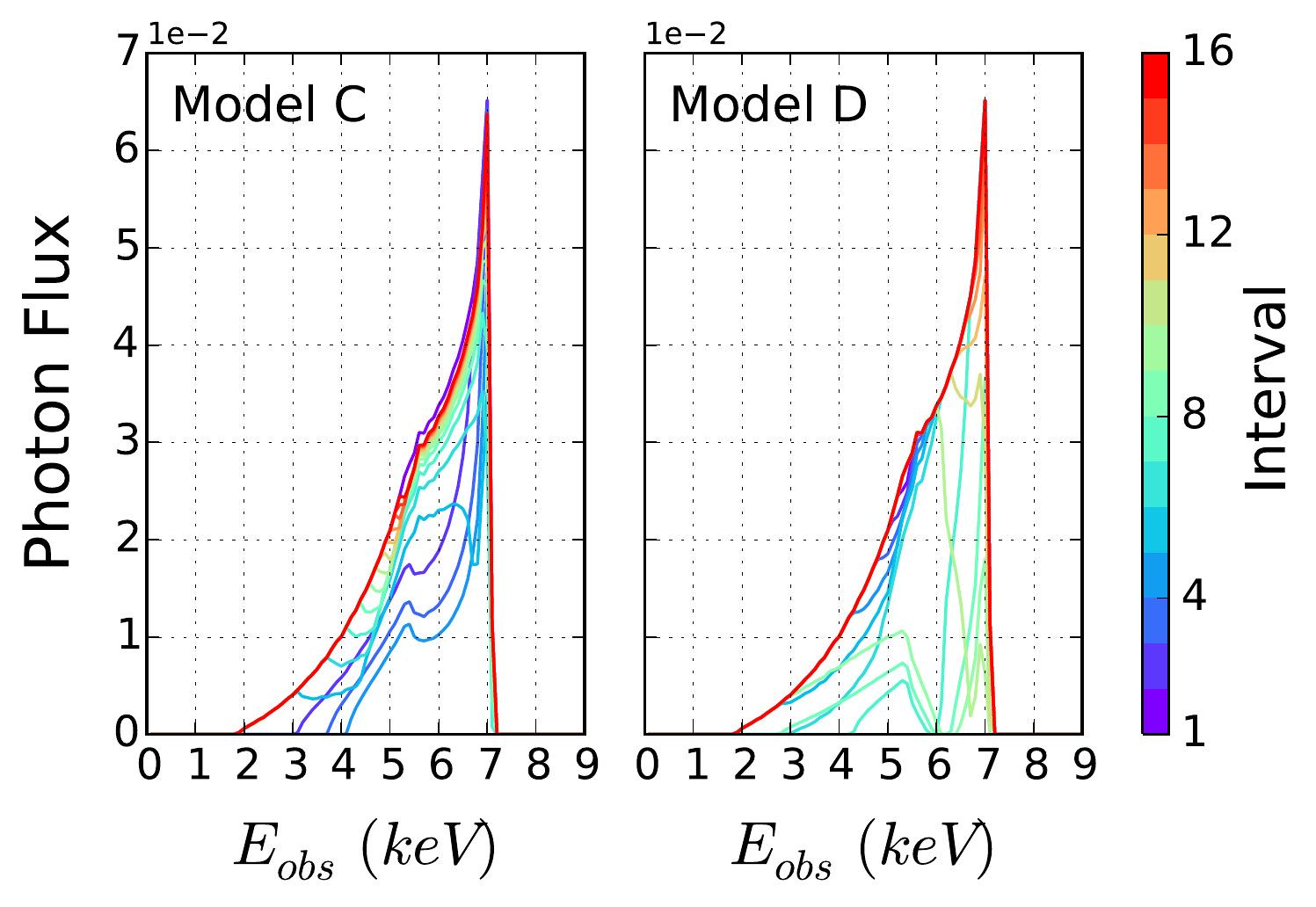}
\end{center}
\vspace{-0.4cm}
\caption{Evolution of the photon flux (left panel) and of the iron line profile (right panel) for the models~C and D. 
See the text for more details. \label{f-cd}}
\vspace{0.6cm}
\begin{center}
\includegraphics[type=pdf,ext=.pdf,read=.pdf,width=10cm]{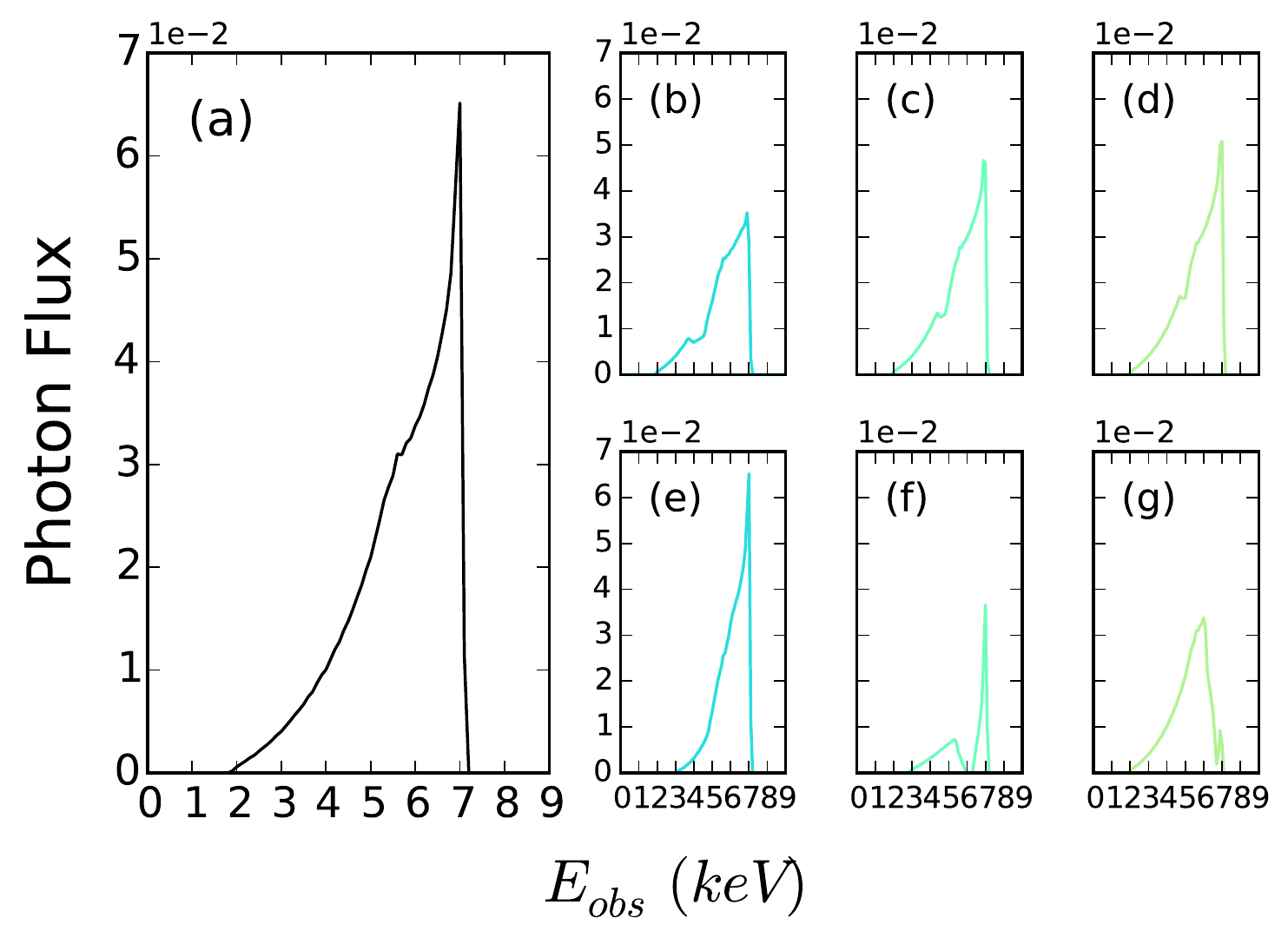}
\end{center}
\vspace{-0.4cm}
\caption{Total iron line profile (panel $a$) and iron line profiles of the model~C (panels $b$, $c$, and $d$) and of 
the 
model~D (panels $e$, $f$, and $g$) at the measurements number~6 (panels $b$ and $e$), 8 (panels $c$ and $f$), and 10 
(panels $d$ and $g$) indicated by the vertical dashed grey lines in the left panel of Fig.~\ref{f-cd}. \label{f-cd2}}
\end{figure}

For the sake of clarity, we show explicitly, in Figs.~\ref{f-cd} and ~\ref{f-cd2}, how the configurations presented in model~C and model~D affect the total flux of the measurement and the iron line profile of a Kerr black hole with spin parameter $a_*' = 0.80$. In particular, Fig.~\ref{f-cd} (left panel) shows specifically the change in the total photon flux, Fig.~\ref{f-cd} (right panel) the change of the iron line profile during the 16 snapshots and Fig.~\ref{f-cd2} just during three selected slices, highlighted by the vertical dashed grey lines in the left panel of Fig.~\ref{f-cd} (left panel).

\begin{figure}
\begin{center}
\includegraphics[type=pdf,ext=.pdf,read=.pdf,width=7.3cm]{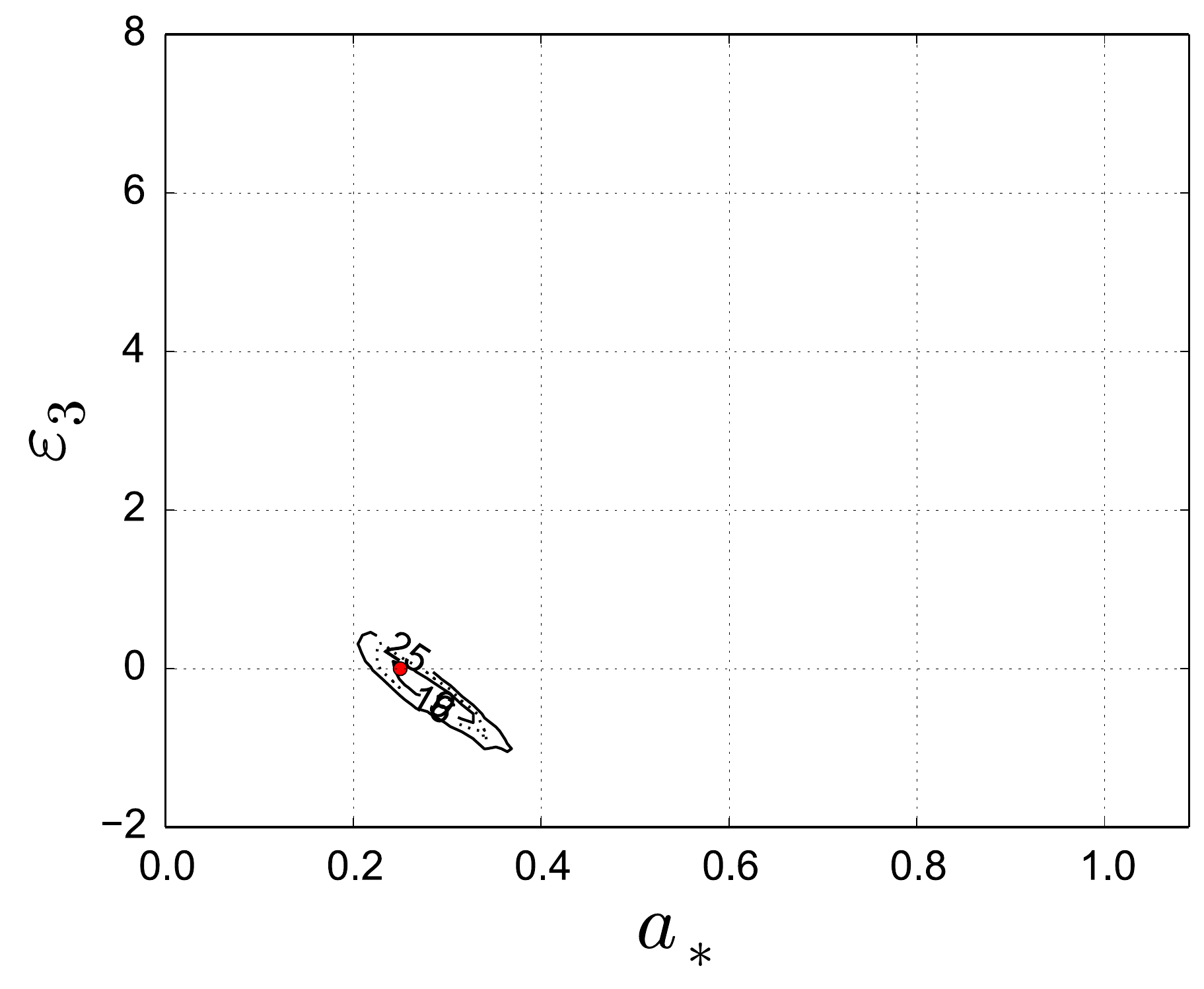}
\hspace{0.5cm}
\includegraphics[type=pdf,ext=.pdf,read=.pdf,width=7.3cm]{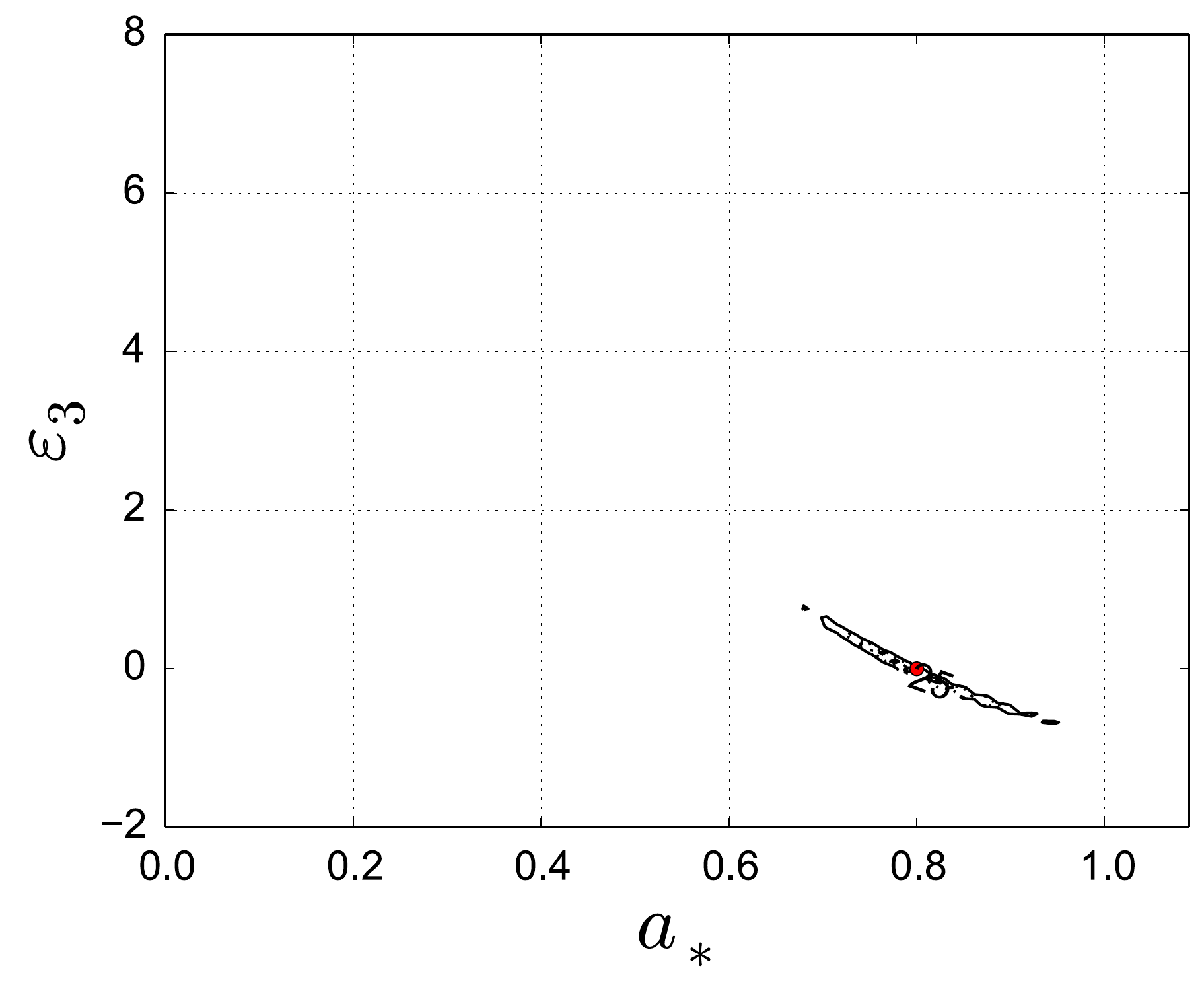} \\
\includegraphics[type=pdf,ext=.pdf,read=.pdf,width=7.3cm]{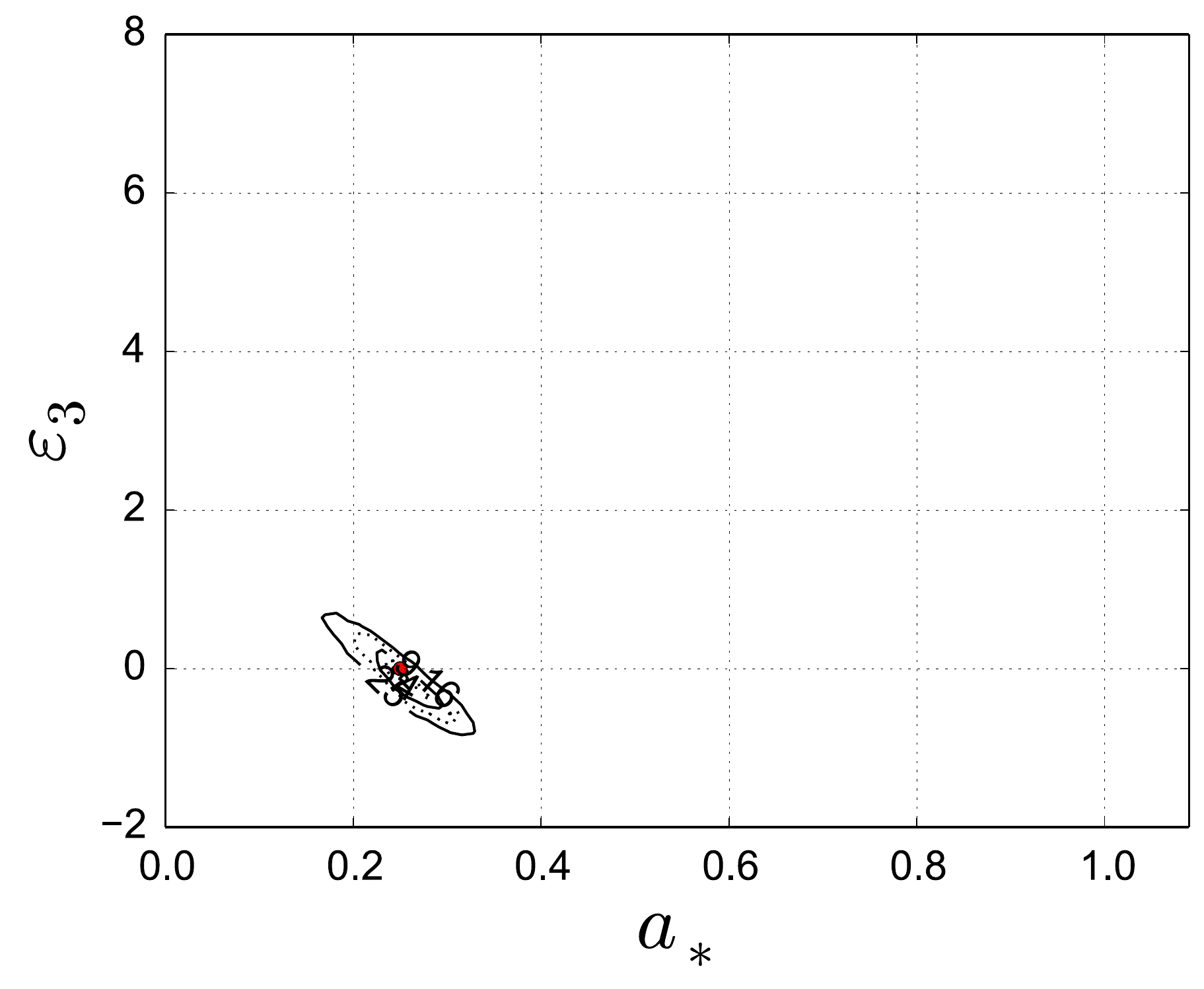}
\hspace{0.5cm}
\includegraphics[type=pdf,ext=.pdf,read=.pdf,width=7.3cm]{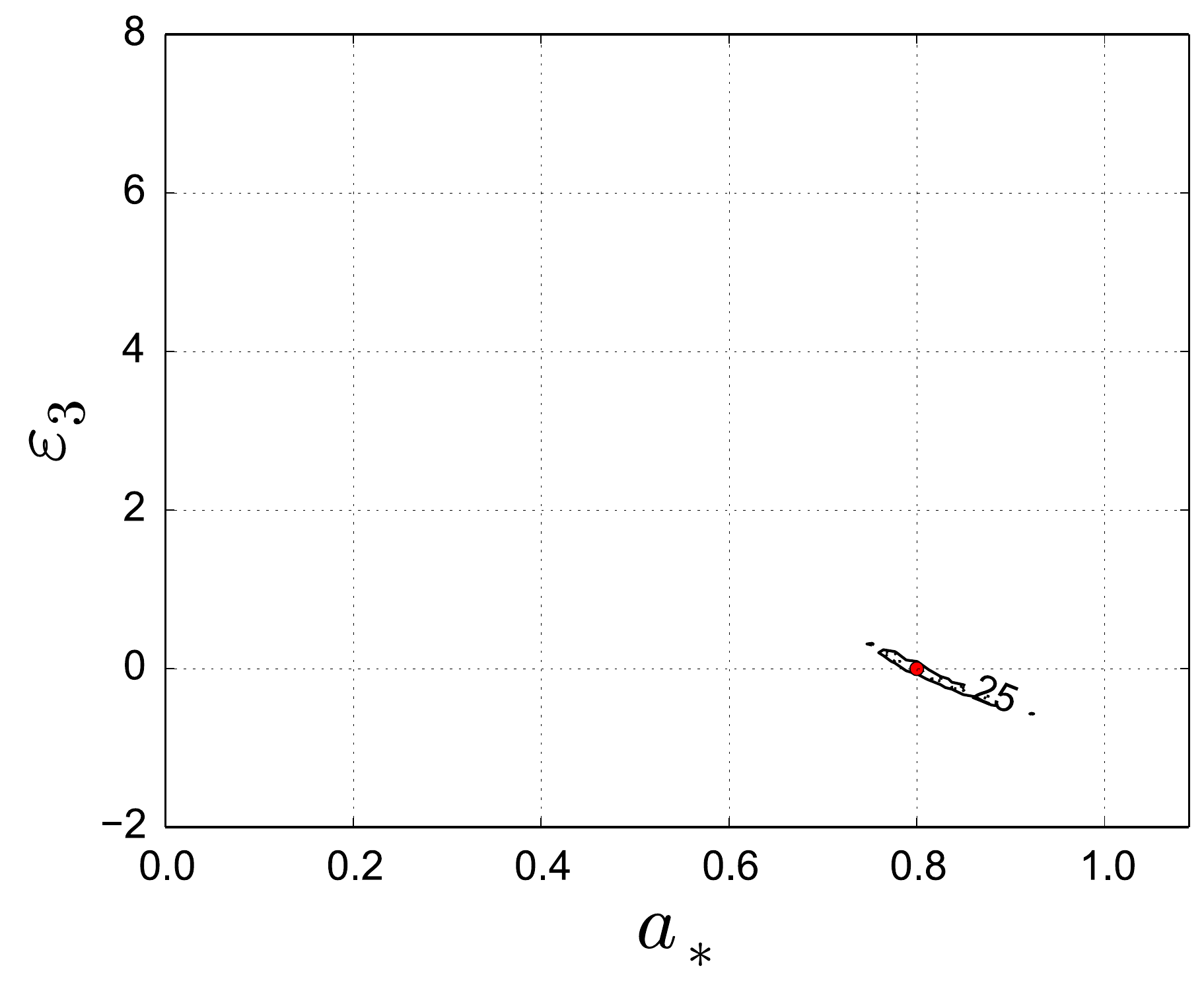} \\
\includegraphics[type=pdf,ext=.pdf,read=.pdf,width=7.3cm]{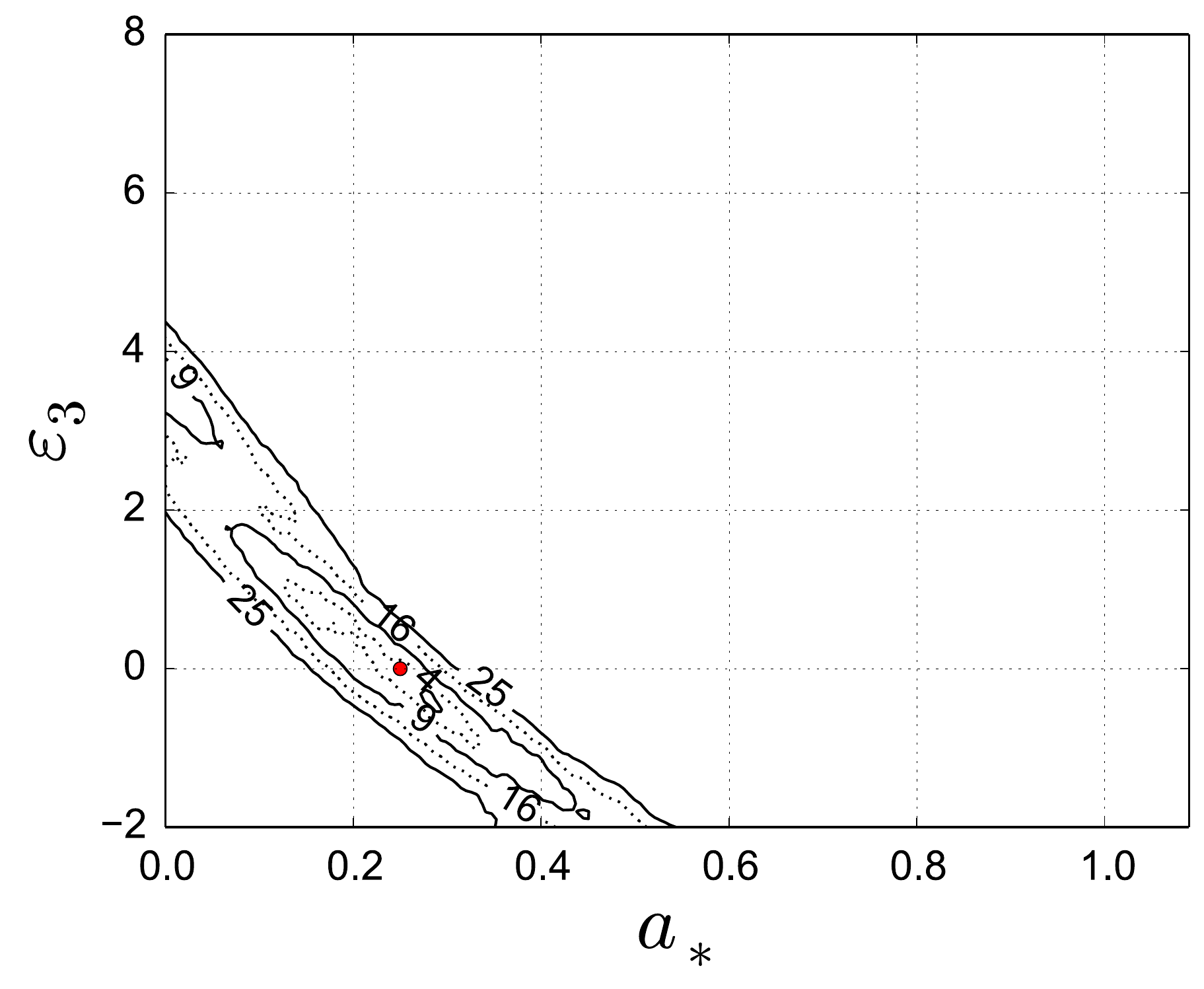}
\hspace{0.5cm}
\includegraphics[type=pdf,ext=.pdf,read=.pdf,width=7.3cm]{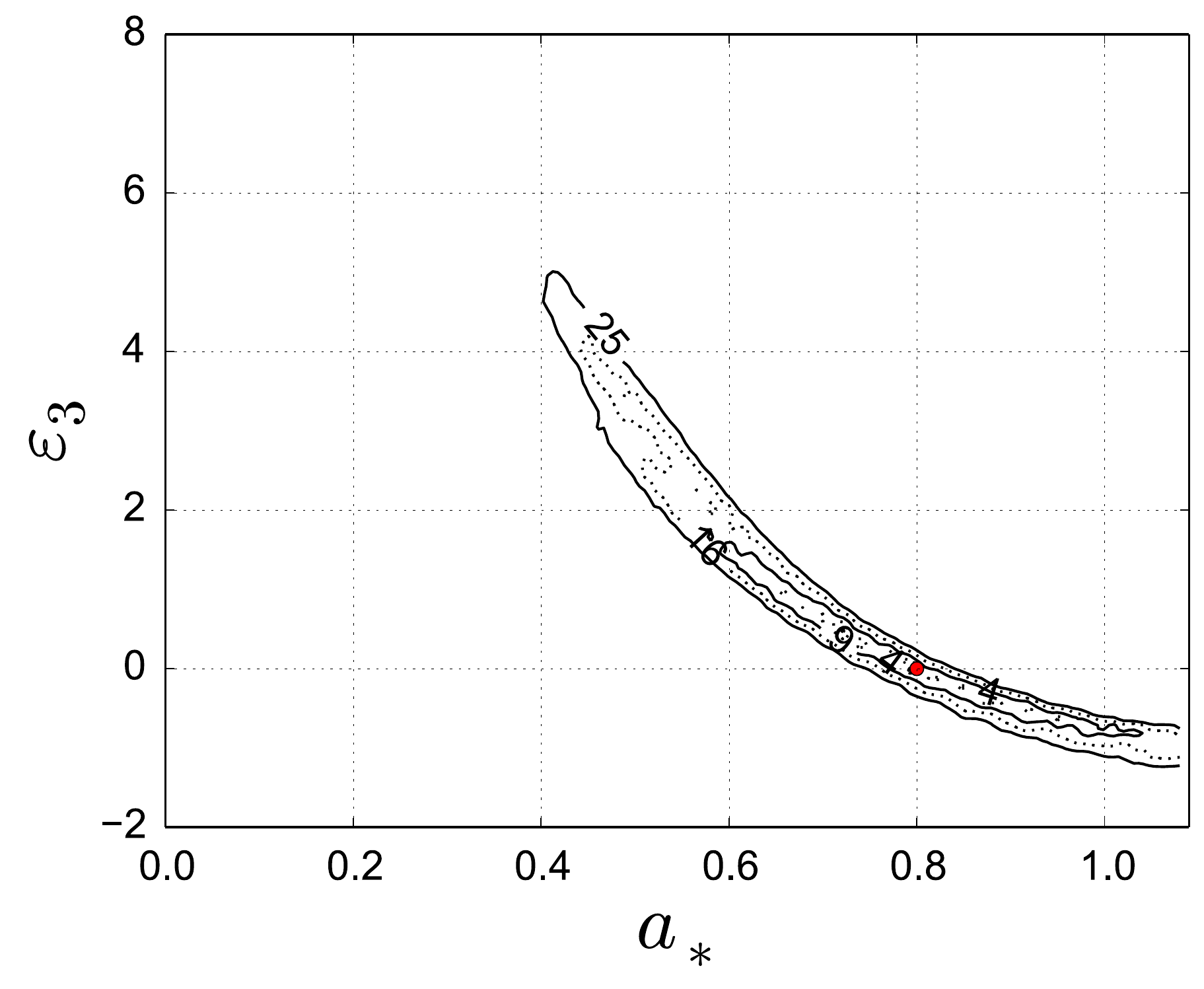}
\end{center}
\vspace{-0.4cm}
\caption{Time-dependent measurement: model~A (top panels), B (middle panels), and C (bottom panels). These plots show 
$\Delta\chi^2$ contours with $N_{\rm line} = 10^4$ (when there is no cloud) from the comparison of the iron line of a 
Kerr black hole simulated using an input parameter $a_*' = 0.25$ (left panels) or $a_*' = 0.8$ (right panels) and a 
viewing angle $i' = 45^\circ$ vs a set of Johannsen-Psaltis black holes with spin parameter $a_*$, non-vanishing 
deformation parameter $\epsilon_3$, and arbitrary viewing angle $i$. See the text for more details. \label{fig4}}
\end{figure}

Fig.~\ref{fig1} already shows the results of the simulations of the model~0 (time-integrated measurement), and Fig.~\ref{fig3} the constraints from the model~D (eclipse measurement). The results of our simulations for the model~A, model~B, and model~C are reported in Fig.~\ref{fig4}, respectively top (model~A), middle (model~B), and bottom (model~C) panels. In the left panels, the reference model is a Kerr black hole with spin parameter $a_*' = 0.25$, in the right panel it is a Kerr black hole with $a_*' = 0.8$. The constraints in the models~A and B remind those found in~\cite{jjc3} for the case of reverberation mapping. The constraints are somewhat stronger than those from a time-integrated measurement with $N_{\rm line} = 10^5$. The constraints become weaker in the model~C, and are similar to those from the model~D in Fig.~\ref{fig3} for an eclipse measurement.

\begin{figure}
\begin{center}
\includegraphics[type=pdf,ext=.pdf,read=.pdf,width=7.3cm]{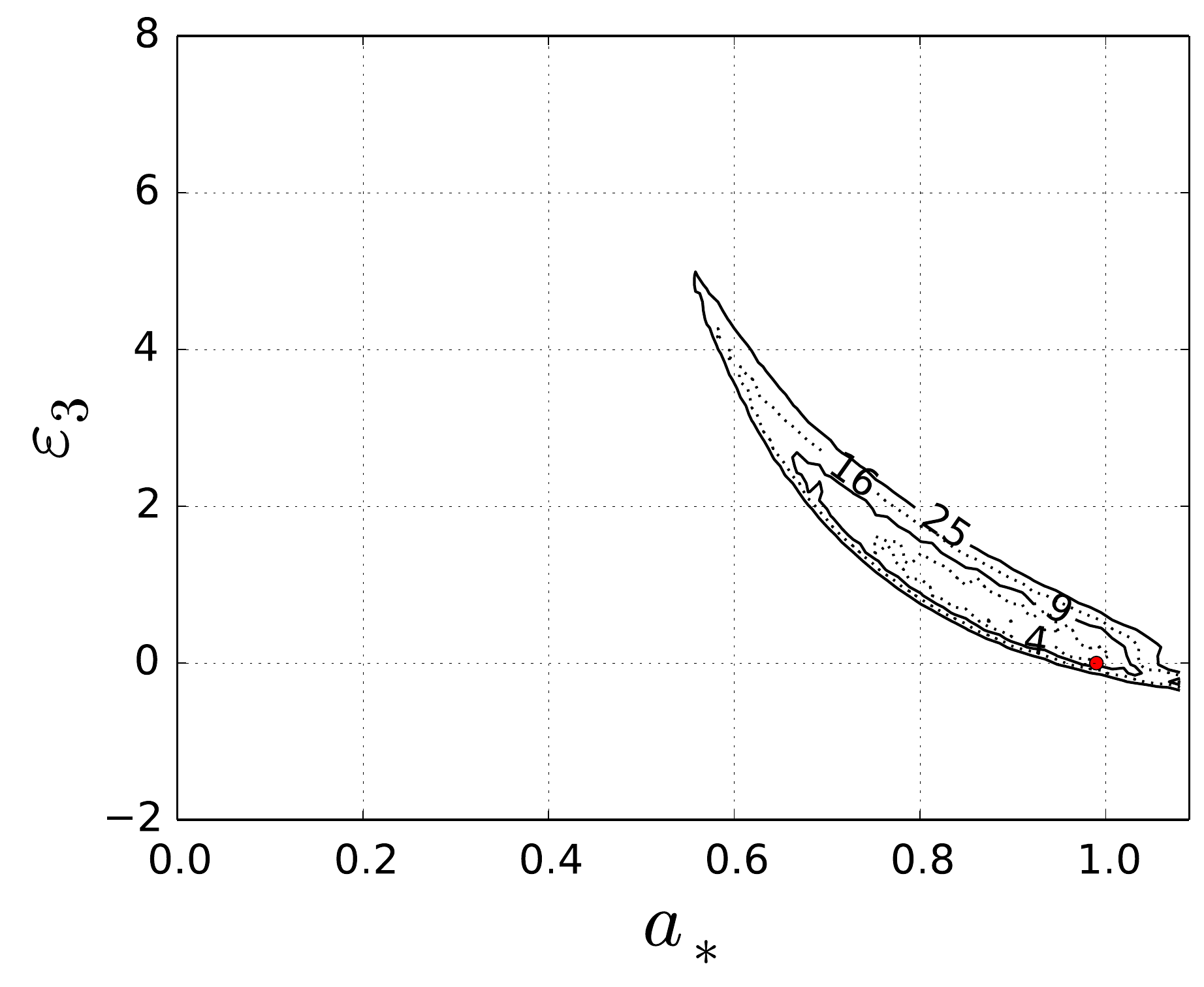}
\hspace{0.5cm}
\includegraphics[type=pdf,ext=.pdf,read=.pdf,width=7.3cm]{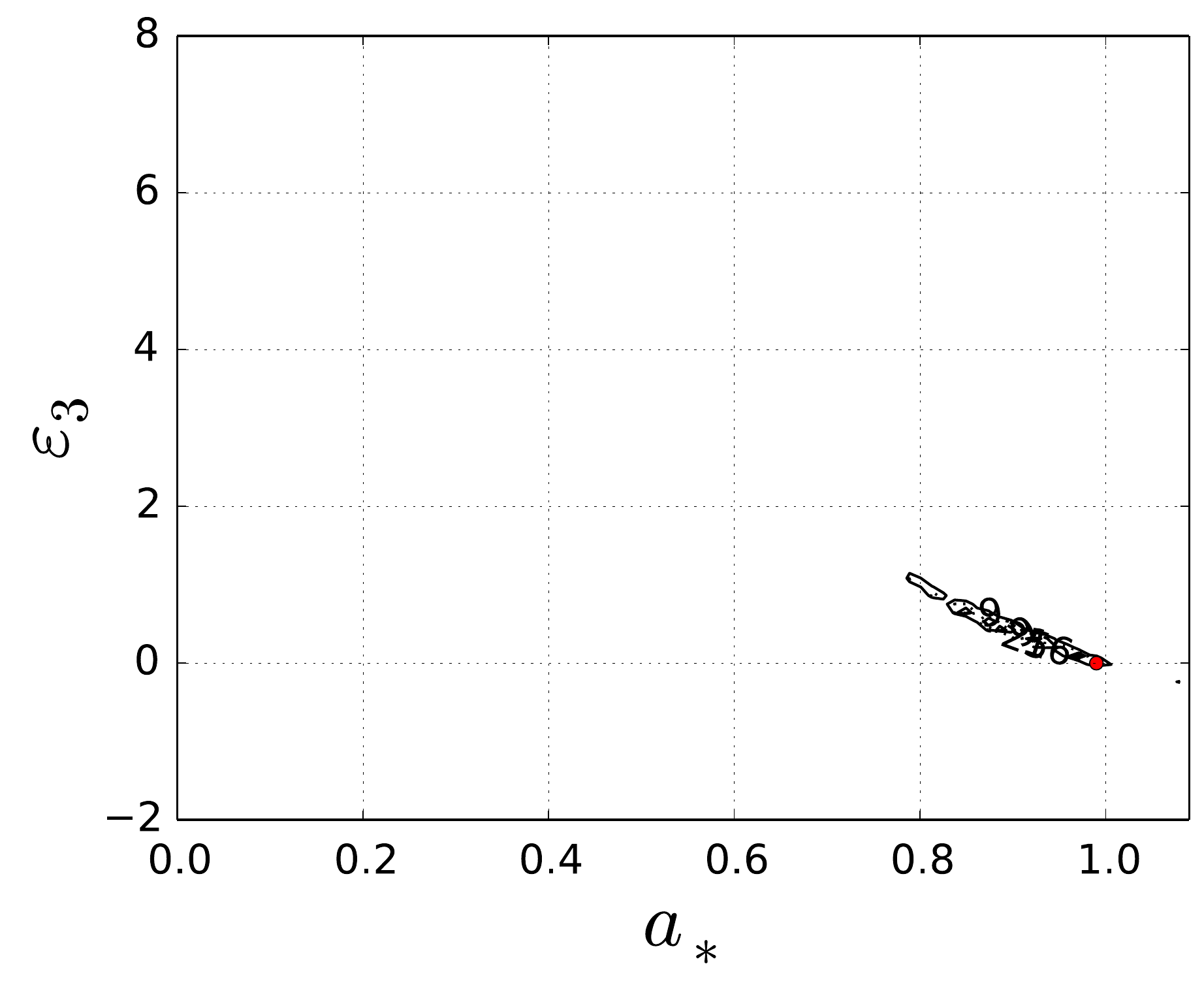} \\
\includegraphics[type=pdf,ext=.pdf,read=.pdf,width=7.3cm]{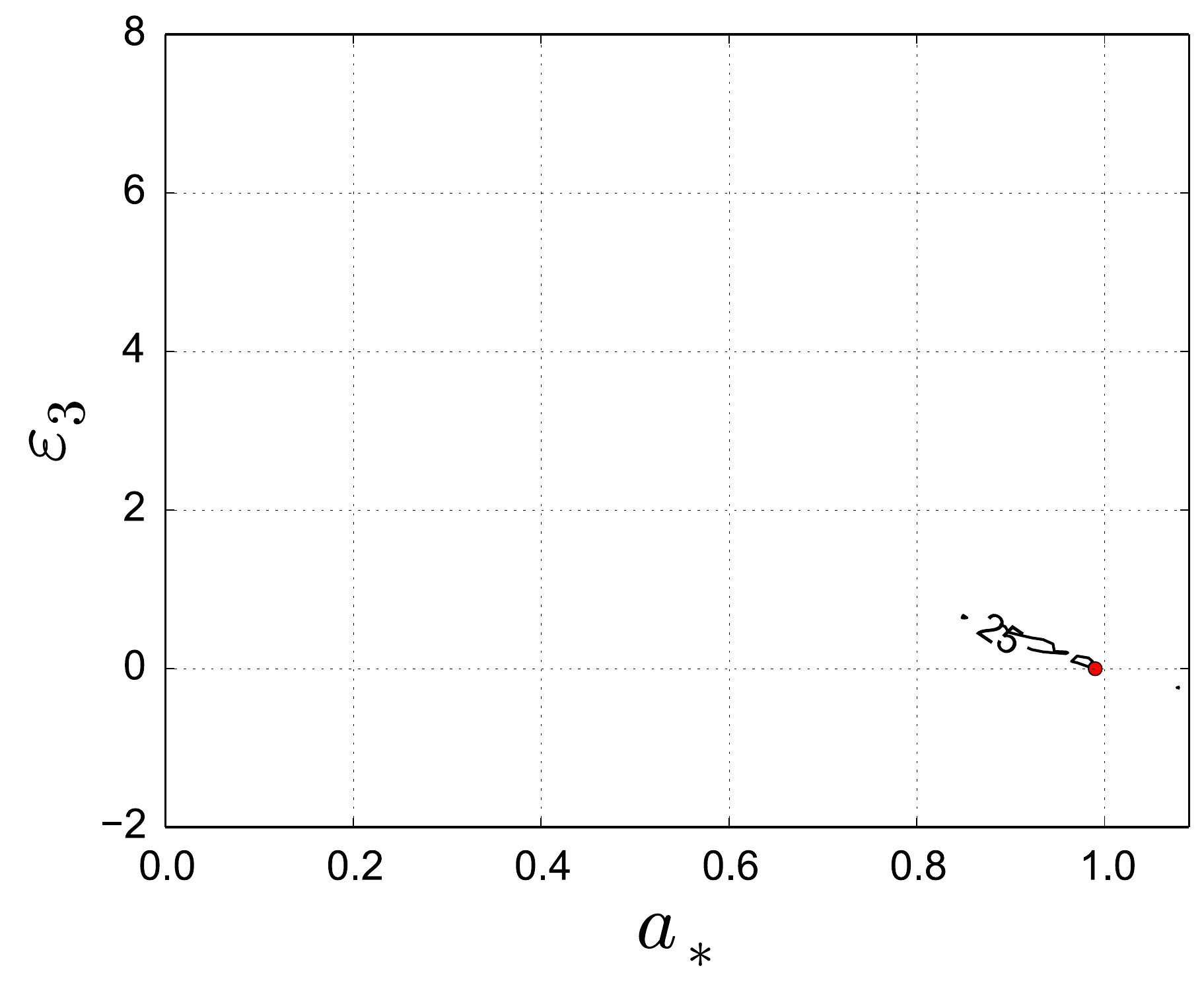}
\hspace{0.5cm}
\includegraphics[type=pdf,ext=.pdf,read=.pdf,width=7.3cm]{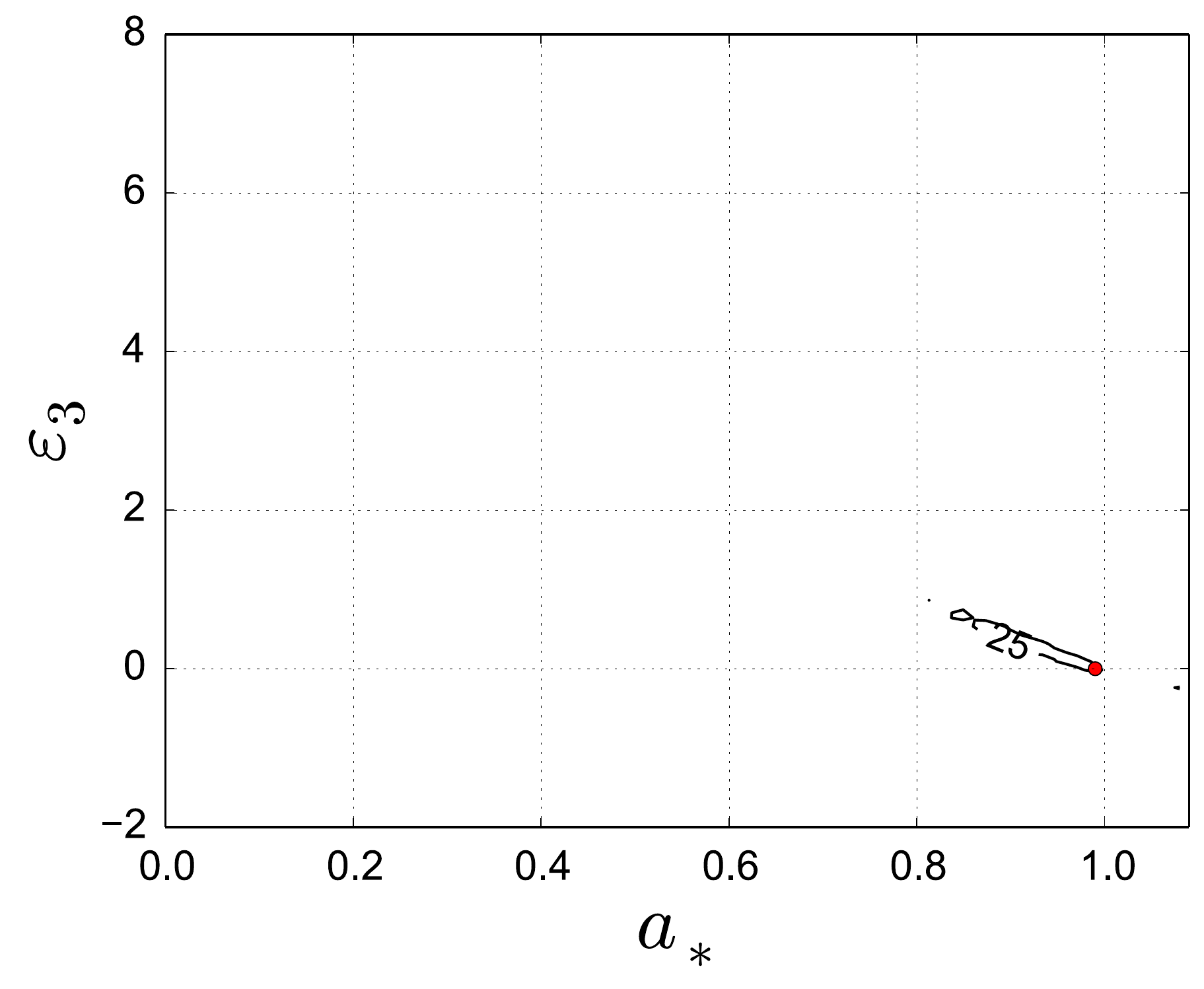} \\
\includegraphics[type=pdf,ext=.pdf,read=.pdf,width=7.3cm]{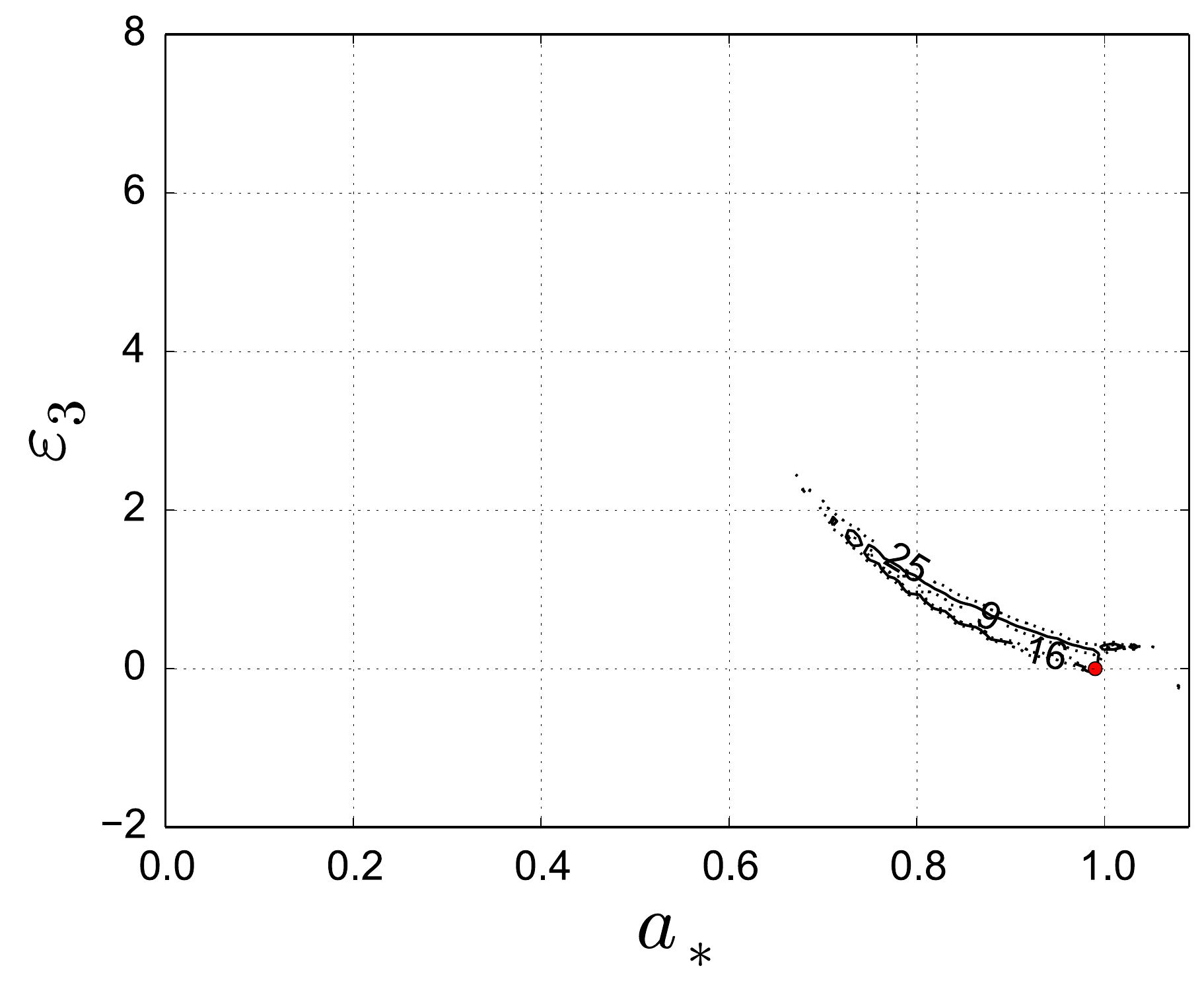}
\hspace{0.5cm}
\includegraphics[type=pdf,ext=.pdf,read=.pdf,width=7.3cm]{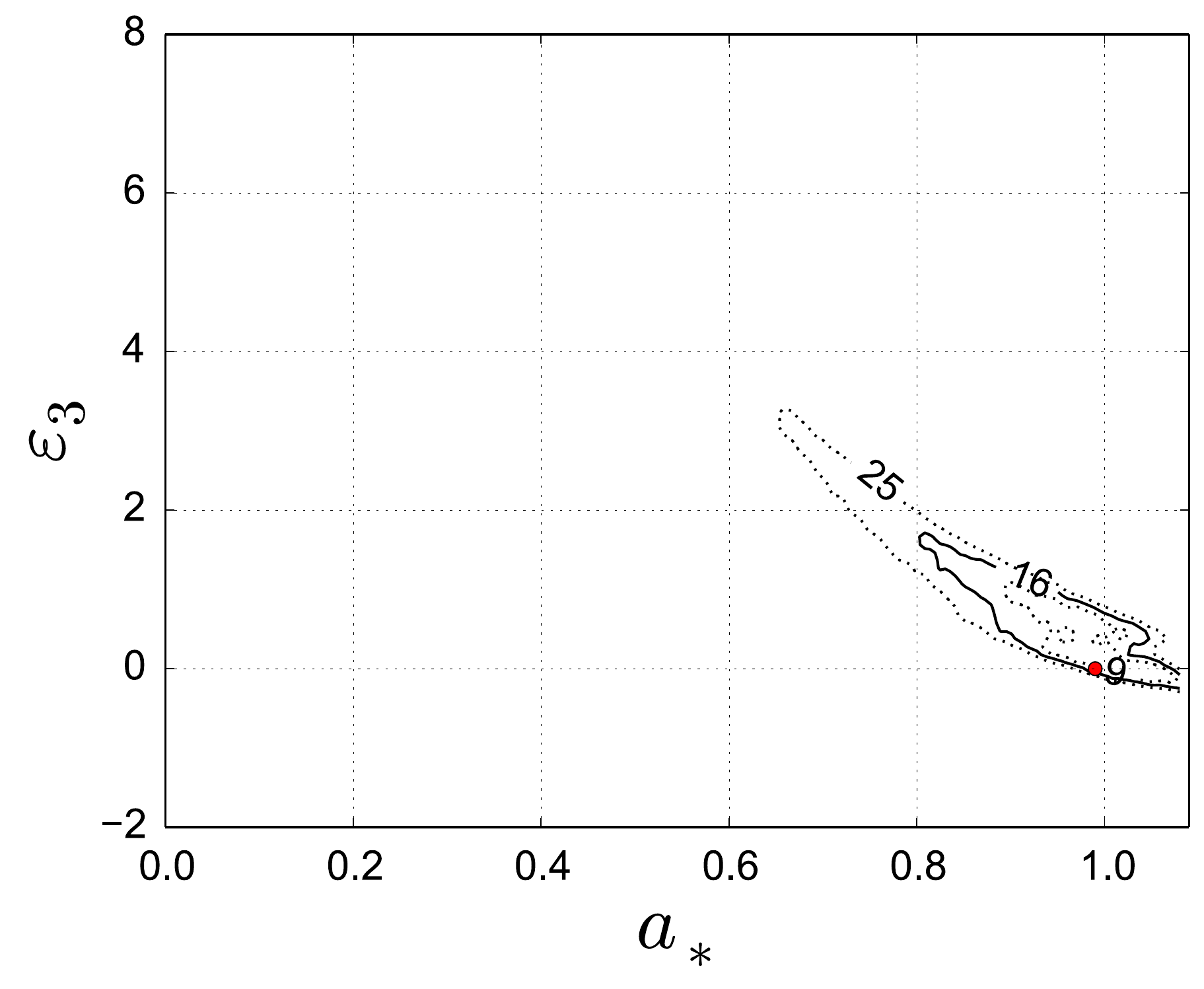}
\end{center}
\vspace{-0.4cm}
\caption{Simulations with a reference Kerr black hole with spin parameter $a_*' = 0.99$ and inclination angle $i' = 45^\circ$: model~0 with $N_{\rm line} = 10^4$ (top left panel), model~0 with $N_{\rm line} = 10^5$ (top right panel),  model~A (central left panel), model~B (central right panel), model~C (bottom left panel), and model~D (bottom right panel). Models A-D are all with $N_{\rm line} = 10^4$ when without cloud. \label{fig5}}
\end{figure}

For completeness, Fig.~\ref{fig5} shows the simulations for a fast-rotating Kerr black hole with spin parameter $a_*' = 0.99$ as reference model. One may indeed think that in this case the relativistic effects are amplified and the eclipse measurement may show some benefits. Actually the picture seems to be similar to the previous cases: the eclipse constraints are somewhat stronger, but they are not like those from reverberation mapping. The constraints from the time-integrated measurement (model~0) are still quite similar to those from the model~C and D, while the constraints from model~A and B with $N_{\rm line} = 10^4$ are comparable, or even something better, than the constraints from the model~0 with $N_{\rm line} = 10^5$.

\section{Conclusions}\label{sec:C}

In Refs.~\cite{jjc1,jjc2,jjc3}, two of us have studied how the iron line in the X-ray spectrum of black hole candidates can be used to test the Kerr metric. We have considered both time-integrated and reverberation measurements. Our results clearly show that the time information in the reverberation measurement can better probe the spacetime geometry around these objects and provide stronger constraints on possible deviations from the Kerr solution.

Motivated by those results, here we have explored another time-dependent measurement: an AGN eclipse in which an obscuring cloud covers different parts of the disk at different times. One may indeed expect to get similar benefits from a reverberation and an eclipse measurements. However, the simple quantitative analysis reported in this work shows that this is not the case. The constraints on the metric that can be obtained from an eclipse observation are typically comparable to those from the standard time-integrated measurement. The exact set-up of the system can somewhat change the results, but we have been unable to find the significant advantages clearly shown in reverberation mapping.

Reverberation and eclipse observations present several differences. However, we have found that the actual difference is made by how the regions of the accretion disk are scanned and one can identify the relativistic effects from each patch of the disk. In the reverberation case, we observe X-ray photons from different regions at different time, and it is easy to reconstruct the spectrum from each region. In an eclipse observation, we have the opposite case, namely we observe the total spectrum of the disk minus the radiation from the region covered by the cloud. We have thus to recover the properties of the covered regions by subtracting iron line profiles with small differences. We have also explored whether our conclusions could change by increasing the photon count in the iron line $N_{\rm line}$. However, we have not been able to find any particular $N_{\rm line}$ where this happens. We really believe that the key-point is how the regions of the accretion disk are scanned.

In this paper, we have adopted the Johannsen-Psaltis parametrisation as a prototype of non-Kerr metric and studied the constraints on $\epsilon_3$. We expect that our main conclusions still hold if we consider other backgrounds. The method discussed here can be applied to any black hole solution, as well as to metrics describing naked singularities, wormhole, etc. We cannot exclude that for some specific spacetime geometry the eclipse measurement provides significant advantages over the time-integrated one, but we can say that the constraining power of this technique is not as promising as reverberation mapping for future tests of the Kerr paradigm.

Lastly, we would like to point out that our results are not in disagreement with the claim in Ref.~\cite{ecl}, in which the conclusion is that an eclipse can help to observe relativistic effects in the iron line. The aim of that paper was to distinguish the scenario in which the iron line signal is produced in the inner part of the accretion disk from the scenario in which the iron line is produced in moving clouds at larger radii. The fact that an eclipse observation can show the Doppler redshifted part and the Doppler blueshifted part of the disk at different times is enough to distinguish the two scenarios. In our case, we already assume that the iron line is produced by fluorescence from the inner part of the accretion disk, and we want to test the Kerr metric.


\begin{acknowledgments}
We would like to thank James Steiner, Laura Brenneman and Martin Elvis for useful comments and suggestions. JJ and CB were supported by the NSFC grants No.~11305038 and No.~U1531117, the Shanghai Municipal Education Commission grant No.~14ZZ001, and the Thousand Young Talents Program. CB also acknowledges also support from the Alexander von Humboldt Foundation. AC-A thanks the Department of Physics at Fudan University, where part of this work was performed, for hospitality during his visit.
\end{acknowledgments}


\end{document}